\documentclass[iop]{emulateapj}%
\usepackage{natbib}
\usepackage{amsmath}
\usepackage{multirow}
\newcommand{\V}[1]{\boldsymbol{#1}}
\DeclareMathOperator*{\argmin}{arg\,min}

\begin{document}

\newcommand{\degree}{\ensuremath{^\circ}}

\title{IMAGING THE ALGOL TRIPLE SYSTEM IN H BAND WITH THE CHARA INTERFEROMETER}


\author{F. Baron\altaffilmark{1}, J. D. Monnier\altaffilmark{1},
 E. Pedretti\altaffilmark{2}, M. Zhao\altaffilmark{3}, G. Schaefer\altaffilmark{4}, R. Parks\altaffilmark{5}, X. Che\altaffilmark{1}, N. Thureau\altaffilmark{6},  T.~A. ten
 Brummelaar\altaffilmark{4}, H.A. McAlister\altaffilmark{4,5}, S.T. Ridgway\altaffilmark{7}, C. Farrington\altaffilmark{4}, J. Sturmann\altaffilmark{4}, L. Sturmann\altaffilmark{4} and N. Turner\altaffilmark{4}} 
 \affil{\altaffilmark{1}Department of Astronomy, University of Michigan, 918 Dennison Building, Ann Arbor, MI 48109-1090, USA}
	\email{fbaron@umich.edu}
 \affil{\altaffilmark{2}European South Observatory, Karl-Schwarzschild-Str. 2, 85748 Garching bei M\"unchen, Germany}
 \affil{\altaffilmark{3}Department of Astronomy, Penn State University, 420 Davey Lab State College, PA 16803, USA}
 \affil{\altaffilmark{4}The CHARA Array of Georgia State University, Mount Wilson CA 91023, USA}
 \affil{\altaffilmark{5}Department of Physics and Astronomy, Georgia State University, Atlanta, GA 30302-3965, USA}
 \affil{\altaffilmark{6}Department of Physics and Astronomy, University of St. Andrews, UK}
 \affil{\altaffilmark{7}National Optical Astronomy Observatory, Tucson, AZ 85726-6732, USA}



\begin{abstract}
Algol ($\beta$~Per) is an extensively studied hierarchical triple system
whose inner pair is a prototype semi-detached binary with mass
transfer occurring from the sub-giant
secondary to the main-sequence primary. We present here the
results of our Algol observations made
between 2006 and 2010 at the CHARA interferometer with the Michigan
Infrared Combiner in the H band. The use of four telescopes with
long baselines allows us to achieve better than~$0.5$ mas
resolution and to unambiguously resolve the three stars. The inner and
outer orbital elements, as well as the angular sizes and mass ratios
for the three components are determined independently from previous
studies. We report a significantly improved orbit for the inner
stellar pair with the consequence of a 15\% change in the primary mass
compared to previous studies. We also determine the mutual inclination
of the orbits to be much closer to perpendicularity than previously
established. State-of-the-art image reconstruction algorithms are used
to image the full triple system. In particular an image sequence of 55
distinct phases of the inner pair orbit is reconstructed, clearly
showing the Roche-lobe-filling secondary revolving around the primary,
with several epochs corresponding to the primary and secondary
eclipses.
\end{abstract}


\keywords{binaries:eclipsing - infrared:stars - stars: imaging - stars: individual (Algol, Bet~Per, HD 19356) - techniques: interferometric - techniques: image processing - {\it Online-only material}: animation, color figures}



\section{Introduction}

Algol ($\beta$~Per, HD 19356) is one of the most extensively studied
stellar systems in the history of astronomy. It is a triple
hierarchical system where an inner eclipsing binary is orbited by a
more distant outer star. The inner system of orbital period 2.87 days
is a prototypical semi-detached binary in nearly circular orbit. The
primary, Algol~A, is a main-sequence star of type B8V. The secondary,
Algol~B, is a subgiant of type K2IV. As Algol~A and Algol~B are thought to
have formed roughly at the same time, one would expect the more
massive star of the pair to become a subgiant first. However this
supposition is contradicted by all observations, showing Algol~B to be
lighter than Algol~A. The solution to this so called ``Algol paradox''
is that most of the original mass from Algol~B, which fills its Roche
lobe, is thought to have been accreted by Algol~A. As the inner pair
is tidally locked, this creates a strong magnetic dynamo effect and
non-thermal emission observed in X-ray and radio. Using radio
interferometry, \citet{Lestrade1993} detected positional displacement
during the orbital revolution of the inner pair and identified Algol~B
as the source of radio emission. In X-rays, giant flares have been
detected on Algol~B \citep{White1986, Schmitt1999,
  Schmitt2003a}. Their sizes were estimated to a few tenths of the Algol~B
radius, and they were shown to follow active and quiescent periodic
cycles of 49 days \citep{Richards2003}. H$\alpha$ profiles provide
evidence for mass transfer through a stream of gas from the Roche lobe
of Algol B to Algol A \citep{Gillet1989}. In parallel, observations
with {\it EXOSAT} \citep{Favata2000}, {\it XMM} \citep{Schmitt2003} and {\it Chandra}
\citep{Drake2003} have shown evidence of an extended corona around
Algol~B. The presence of a large, permanent coronal loop around
Algol~B, oriented toward Algol~A, has very recently been confirmed by
radio images from the HSA \citep{Peterson2010}. 

The tertiary component, Algol~C, was first discovered by radial
velocity measurement \citep{Curtiss1908} and revolves around the inner
pair over a period of about 680 days without eclipsing. It is an Am
star with metal and hydrogen lines presenting F1V characteristics
\citep{Richards1993}. Earlier studies based on astrometric,
photometric, and radial velocity data allowed the determination of the
masses of the inner stars \citep{Hill1971, Tomkin1978} and of Algol~C
\citep{Bachmann1975}. Then light curve meta-analyses combining
previous visible, infrared and ultraviolet observations
\citep{Soderhjelm1980, Richards1988} produced estimates of the orbital
elements for the inner and outer orbits, as well as for the surface
temperatures, $\log g$, radii and masses of the three stars. With the
advent of speckle interferometry, it became possible to spatially
resolve the outer component \citep{Labeyrie1974, McAlister1979},
leading to the determination of orbital parameters for the outer orbit
\citep{Bonneau1979}. Optical interferometry then improved upon these
results with the Mark III interferometer \citep{Pan1993}. However, due
to $180 \degree$ ambiguities in early speckle observations, all
studies based on speckle results \citep{Bonneau1979, Soderhjelm1980,
  Pan1993} produced ascending node estimates for the inner or outer
orbits in disagreement with the polarimetric \citep{Rudy1978} and
radio \citep{Lestrade1993} observations. Therefore, as for given
inclinations, the angle between the inner and outer orbits solely
depends on the difference of the ascending nodes, this ambiguity
implied that the orbits were found either roughly perpendicular or
coplanar. As coplanarity was ruled out by the shallow eclipse depth of
photometric observations \citep{Soderhjelm1980}, ad-hoc corrections
had to be made to the ascending nodes of speckle-based studies. More
recently, \citet{Csizmadia2009} resolved the inner pair with long
baseline interferometry at the Center for High Angular Resolution
Astronomy (CHARA), using the CLASSIC instrument working in the near
infrared $K_s$ band, in combination with {\it Very Long Baseline
  Interferometer} radio observations. They determined the orbit of the
inner binary to be in prograde rotation, in disagreement with the
retrograde movement found previously in radio by
\citet{Lestrade1993}. \citet{Zavala2010} recently solved these
long-standing issues, by simultaneously resolving all three stellar
components in the optical with the Navy Optical Interferometer
(NOI). Using angular referencing, the most precise inner and outer
orbits to date were derived, with positional errors on the stellar
components of the order of a milli-arcsecond. The outer orbit was
shown to be prograde and the inner orbit retrograde. The latest {\it
  Very Long Baseline Array} radio measurements are in agreement with
this orientation \citep{Peterson2011}. \citet{Zavala2010} also
presented the first reconstructed interferometric image of all three
stars, though the short baseline lengths used at NOI precluded full
separation of the inner pair components.

In this paper, we present the results of our Algol observations with
the Michigan InfraRed Combiner (MIRC) in the H band, making use of the
long baselines available at CHARA. In Section~\ref{model_fitting}, the
relative positions of the three components are determined by
model-fitting for the 55 epochs of our data set. New estimates of the
orbital parameters are then derived in Section~\ref{orbital_solution}
using these stellar positions. Finally, images of the inner orbit are
reconstructed for each epoch in Section~\ref{image_reconstruction},
where we discuss the potential evidence for the phenomena detected by
radio and X-ray instruments.

\section{Observations and data reduction}\label{reduction}

Our observations of Algol were conducted between 2006 and 2010 at the
Georgia State University CHARA Array interferometer using the MIRC
instrument. The CHARA Array, located on Mount Wilson, CA, consists of
six 1~m telescopes and is the largest optical/IR interferometer
array in the world \citep{Brummelaar2005}. Its 15 baselines range from
34~m to 331~m, providing resolutions up to $\sim 0.5$~mas at H band and
$\sim 0.7$~mas at K band. Thus the data collected constitute the
highest angular resolution data available to date on Algol. MIRC was
used to combine four CHARA telescopes together for true
interferometric imaging in H band, providing six visibilities, four closure
phases and four triple amplitudes simultaneously in eight narrow spectral
bands for each snapshot \citep{Monnier2006}.

\begin{figure}
\centering
\plottwo{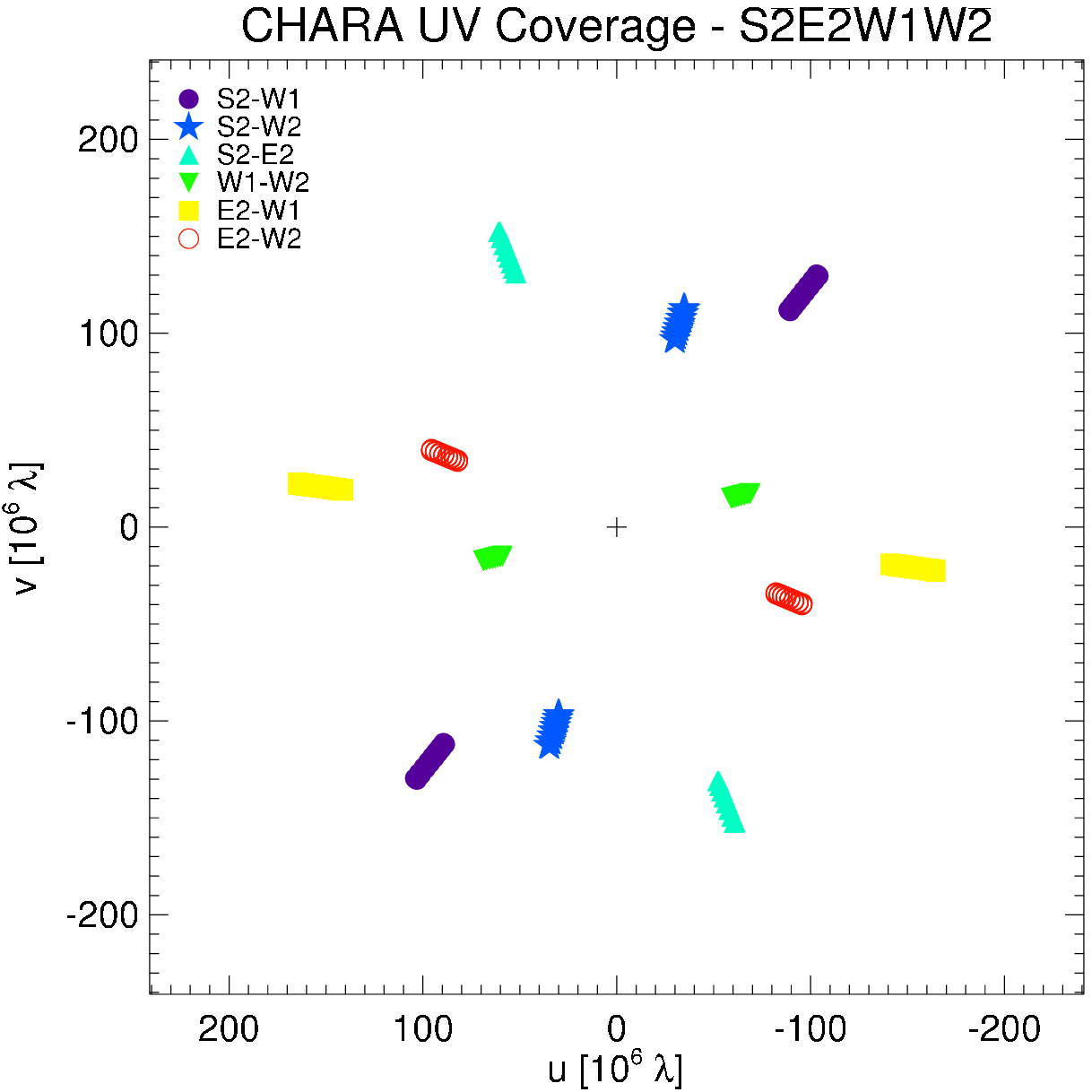}{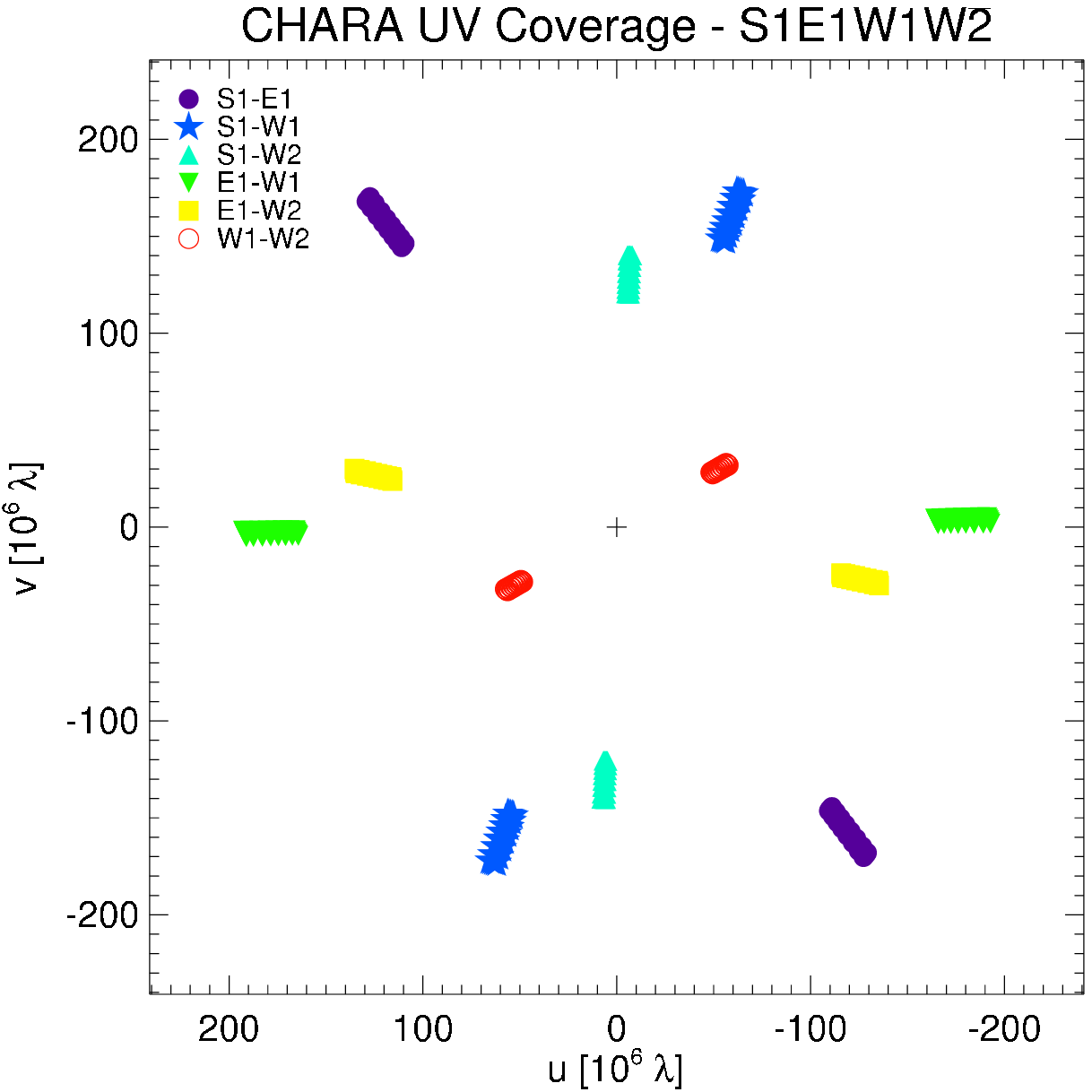}
\caption{Typical {\it uv} coverage for one of our Algol ``split'' data
 set when using the telescope configuration S2-E2-W1-W2 (left) and
 S1-E1-W1-W2 (right). \label{fig-uv}}
\end{figure}

Our observations total 20 nights, making the accumulated
interferometric data on Algol one of the largest datasets to date at
CHARA. In particular, the data covers most phases of the inner pair
eclipse. A complete observing log is listed in Table~\ref{tbl-1},
including our calibrators. The spacing of Algol observations is not
regular: Algol was sometimes observed only as a backup target, when
atmospheric conditions did not allow for other fainter targets. The
duration of most of these observations generally did not
exceed~1~h. Two array configurations were used, S1-E1-W1-W2 and
S2-E2-W1-W2. The full complement of fringes was always obtained except
on 2009 August 18 when bad seeing prevented the acquisition of the E1
fringes. Both configurations are well-adapted to imaging, with roughly
equal Fourier coverage in all directions. Figure~\ref{fig-uv} presents
the typical Fourier coverage achieved within one of our final data
sets.

The standard observing procedures were followed, and the data were
reduced using the standard MIRC pipeline described in
\citet{Monnier2007}. After frame co-adding, background subtraction and
a Fourier transform of the raw data, fringe amplitudes and phases are
used to form squared-visibilities and triple products. On all data
prior to 2010, the photometric calibrations are estimated using
shutter matrix measurements and partial beam chopping. Starting in
2010 data, the calibration relies on the newly available (and far
superior) photometric channel technique, allowing flux measurement in
parallel with the fringes \citep{Che2010}. During the next step,
calibrators are used to calibrate the drifts in overall system
response. Table~\ref{tbl-2} presents the sizes of the calibrators, as
estimated from our ongoing study of ``good'' interferometric
calibrators and based on three independent photometry methods,
adopting different color--magnitude relations to compute the stellar
sizes. It has to be noted here that 37 And is currently suspected to
be a binary, though with a flux ratio between components large enough
not to significantly affect the calibration. After calibration, the
pipeline outputs the final calibrated power spectra and bispectra into
files compliant with the OIFITS standard \citep{Pauls2005}.

Based on the orbital parameters obtained by previous studies, we
expect significant stellar movement on short timescales. The orbital
period of the inner pair is about 2.87 days, meaning that Algol~B is
moving relatively to Algol~A by about $0.07$~mas in an
hour. The movement of Algol~C also impacts the visibilities through
high order lobes which oscillate rapidly with its separation from the
inner pair (ranging from 15 to 70 mas), though this phenomenon is
somewhat mitigated by its much longer orbital period (about 680 days).

\begin{figure}
\centering
\plotone{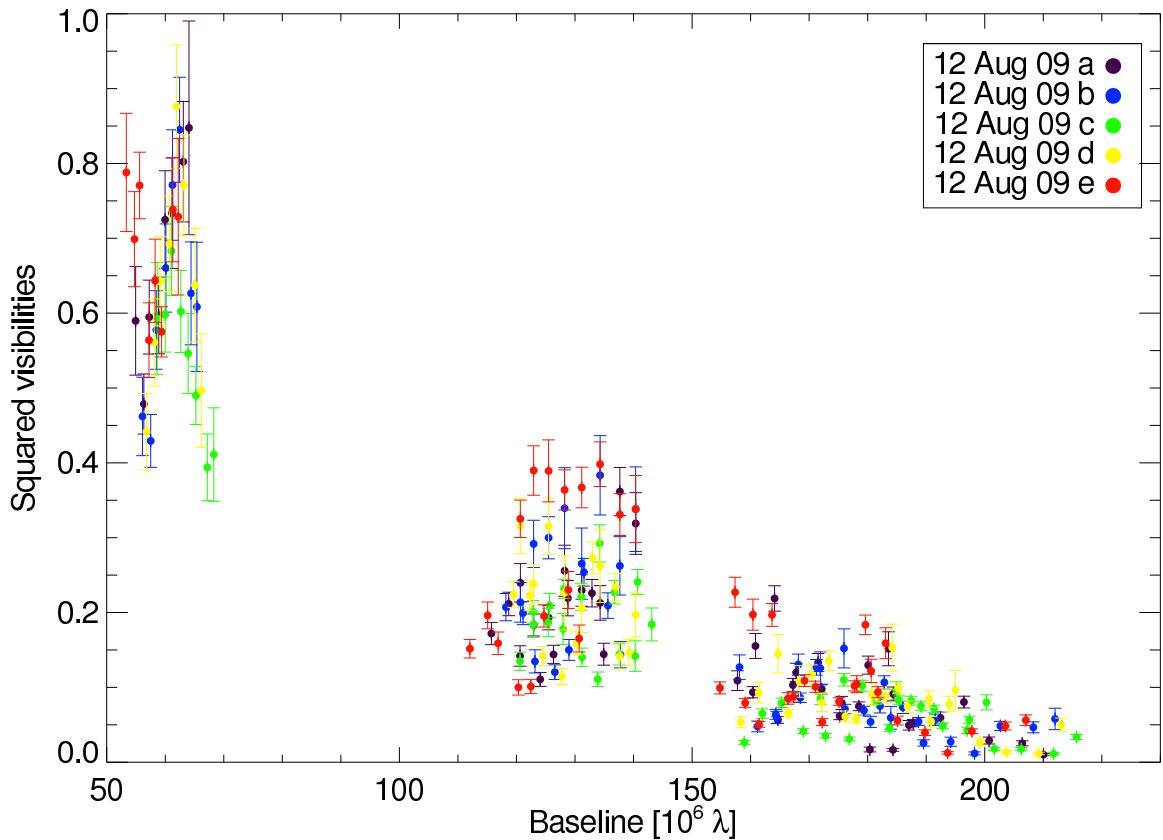}
\caption{Squared visibilities collected on 2009 August 12,
  demonstrating how the dataset has been splitted into 5 epochs (2009
  August 12 a,b, \ldots e). Both Algol visibilities and the {\it uv}
  coverage are significantly changing over the course of
  20~minutes. For clarity we only show one--fifth of the full
  data set.\label{fig-splitting1}}
\end{figure}

\begin{figure}
\centering
\plotone{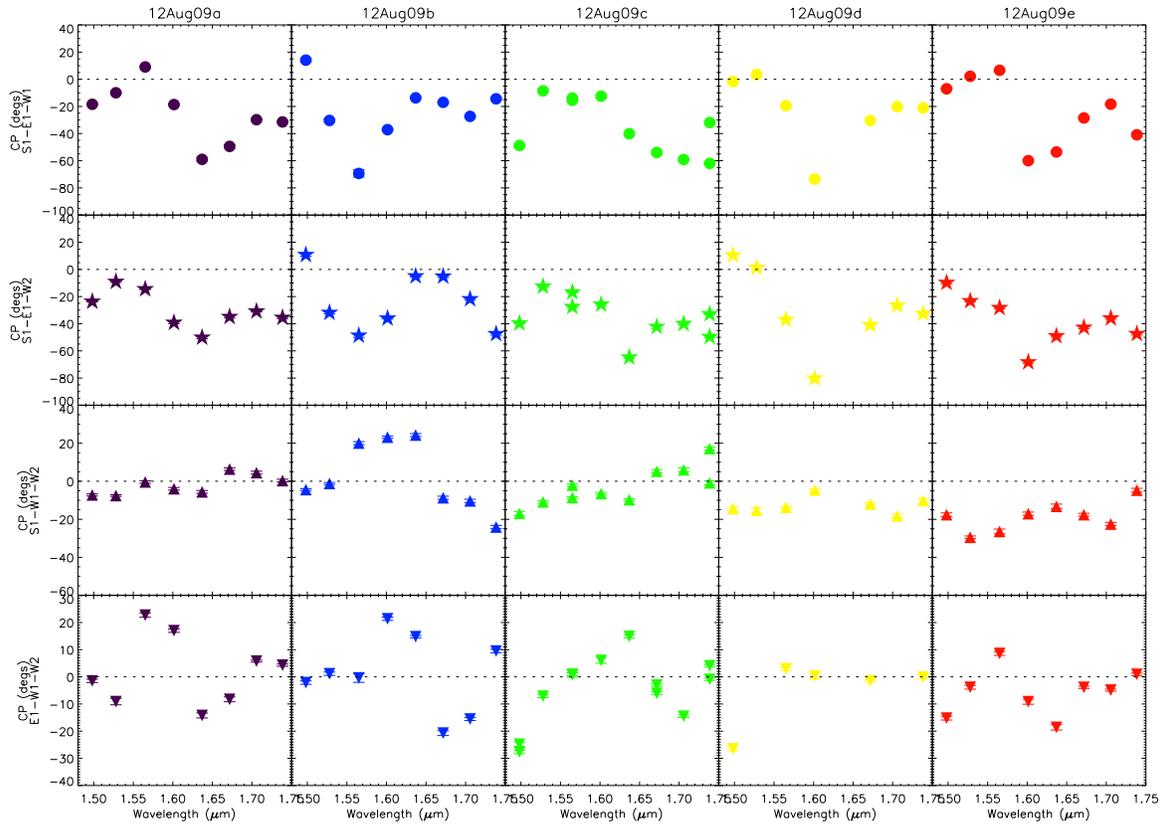}
\caption{Closure phase data collected on 2009 August 12 presented as a
  function of wavelength and sorted by epochs. The error bars on the
  closure phases are of the order of~$1\degree$.\label{fig-splitting2}}
\end{figure}

While model-fitting can possibly be made to account for these time
dependencies, image reconstruction packages currently cannot. At best,
an image reconstructed using our data accumulated over an hour would
be blurred, and at worse it would not converge at all toward a
meaningful solution. Thus, the data have to be split into shorter
sequences of no more than 10 minutes, finding a compromise between
reasonable Fourier coverage and temporal consistency. This splitting
is applied to about 60\% of the nights and results in a total of 55
different files, each of which contains about 240 power spectra and
160 bispectra, sufficient for imaging without
blurring. Figures~\ref{fig-splitting1} and~\ref{fig-splitting2}
illustrate how the data accumulated on 2009 August 12 have been split
into chunks.

\section{Model fitting}\label{model_fitting}

The purpose of model-fitting here is to derive the position, shape and
flux parameters of all three Algol components. These parameters will
then be used to determine the orbital parameters of the inner orbit
(noted A--B) and outer orbit (noted AB--C), as well as to generate
realistic prior images for later image reconstruction. Our general
procedure is to minimize the residuals between the interferometric
observables derived from our current model and the corresponding
interferometric data, using a Levenberg-Marquardt algorithm. Two
possible paths were initially examined: either directly fit the
orbital parameters on the full data set, or estimate a separate set of
parameters for every single epoch and then fit the orbital parameters
using these positions. As Algol~B is non-spherical (because it is thought to
fill its Roche lobe), the two-dimensional projection of its surface onto the
observing plane varies from night to night \citep{Richards1992}. The ``global
fit'' approach would require a complex three--dimensional model of the three
stars, while the ``night-to-night fit'' approach we finally settled for is more
flexible, as well as more manageable in terms of computing power and memory
requirements.

Each star is modeled here as an ellipse of uniform brightness
distribution. As the angular sizes of the components are only few
times the array resolution, the use of limb-darkening coefficients
would have a nearly negligible impact on the visibilities. In fact,
limb-darkened models are not strictly necessary for orbit fitting and
even detrimental as they increase the number of parameters to
estimate. Furthermore we expect image reconstruction to retrieve the
true flux distribution. A major effect to take into account when
modeling the system is the bandwidth smearing effect on Algol~C,
i.e. Algol~C's contribution to the visibilities is affected by its
large separation from the delay-tracking center (Algol~A) and the
width of each of the MIRC spectral channels. With this effect, the
total number of parameters to estimate for each epoch is $23$: the
principal axes and angles for Algol~B and Algol~C relative to Algol~A
(the available interferometric data does not give access to absolute
positions), the relative flux contributions of Algol~A and Algol~B to
the total flux (normalized to unity), the shape parameters of the
three elliptical disks (semimajor axis, ellipticity, and angular
orientation), and the bandwidth smearing factors for Algol~C (one for
each spectral channel).

The~$\chi^2$ formed from power spectra and bispectra is known to be
multimodal: as the Levenberg--Marquardt is a gradient algorithm, the
parameters have to be initialized close to their true value to prevent
falling into local minima. Attempting to estimate the $23$~unknowns
simultaneously would only lead to inconsistent results. We determined
that the positions of Algol~B and Algol~C relative to Algol~A
constitute the parameters with the most influence on the
$\chi^2$. Therefore our solution to this convergence problem involves
simultaneous systematic grid searches on both Algol~B and Algol~C
Cartesian coordinates (relative to Algol~A), coupled to
Levenberg-Marquardt optimizations on the remaining parameters. In
addition, some parameters can safely be fixed, based on the literature
of the expected physical properties of the system. As Algol~A and B
are thought to be tidally locked \citep{Richards1992}, the inner orbit
is posed circular. The principal axis of Algol~B is supposed to be
along the direction A-B, as expected for a Roche-lobe filling
secondary. Algol~A and Algol~C are both assumed perfectly circular,
because they are not rapid rotators and are not filling their Roche
lobes. With these assumptions the $\chi^2$ fit on the remaining number
of parameters is then well constrained as each data set contains
between 120 and 240 power spectra as well as between 60 and 160
bispectra (triple amplitudes and closure phases). Out of the eclipse
periods, lower and higher bounds are imposed on each parameter solely
to enforce the physical meaning of parameters. During the eclipses,
the aspect ratios of Algol~B are kept small ($<1.05$).

The fitting is done in two steps -- first a coarse grid search on
Algol~B and Algol~C positions, then a finer one. The coarse search
uses a $10 \times 10$ grid with spacing $0.1$~mas aimed at determining
the approximate positions of the components. For each epoch, the
center of the Algol~C grid is initialized at its expected position
based on the orbital elements from the NOI results \citep{Zavala2010},
while the center of the Algol~B grid search is simply set to the
origin $(0,0)$ (i.e., Algol~A position). For a given night, this setup
of the coarse grid search covers all the probable positions of the
components. Then the finer grid search refines the estimates, using a
$10 \times 10$ grid with spacing $0.02$~mas around the most likely
positions found during the coarse search. To insure convergence during
both the coarse and fine searches, the fluxes and angular diameters of
the star are always optimized first, while the secondary ellipticity
and the bandwidth smearing factors are kept fixed to their initial
estimates; then the constraints on these are relaxed and all
parameters are re-optimized.

To compute the visibility contribution of Algol~C, an analytic
expression of the complex visibilities of a bandwidth-smeared ellipse
is directly employed. For the visibility contribution of the inner
pair, the presence of epochs where one component eclipses the other
prevents the use of analytic expressions. Algol~A and Algol~B are therefore
modeled as ellipses on an image array of $64$ by $64$ pixels with a
pixel size of $0.05$ mas. To prevent problems linked to the
discretization of the ellipse sizes with this pixellation, the edges
of the ellipses are smoothed by a sigmoid function, so that the
$\chi^2$ varies continuously with the size parameters. The complex
visibilities are then computed by applying a Discrete Fourier
Transform to the image. Finally the model power spectra and bispectra
are derived from the visibilities, and the reduced $\chi^2$ is
computed.

\begin{figure}
\centering
\plottwo{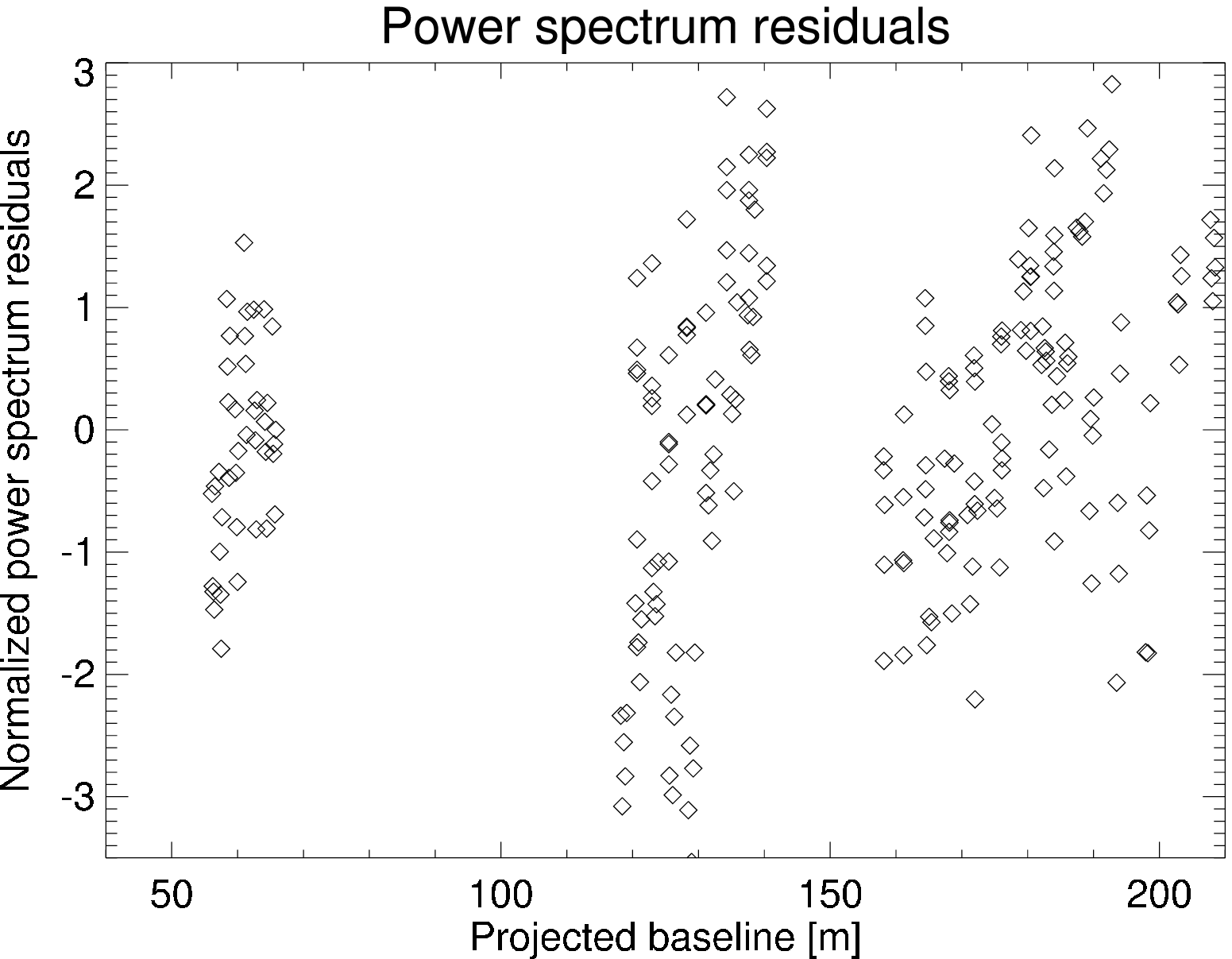}{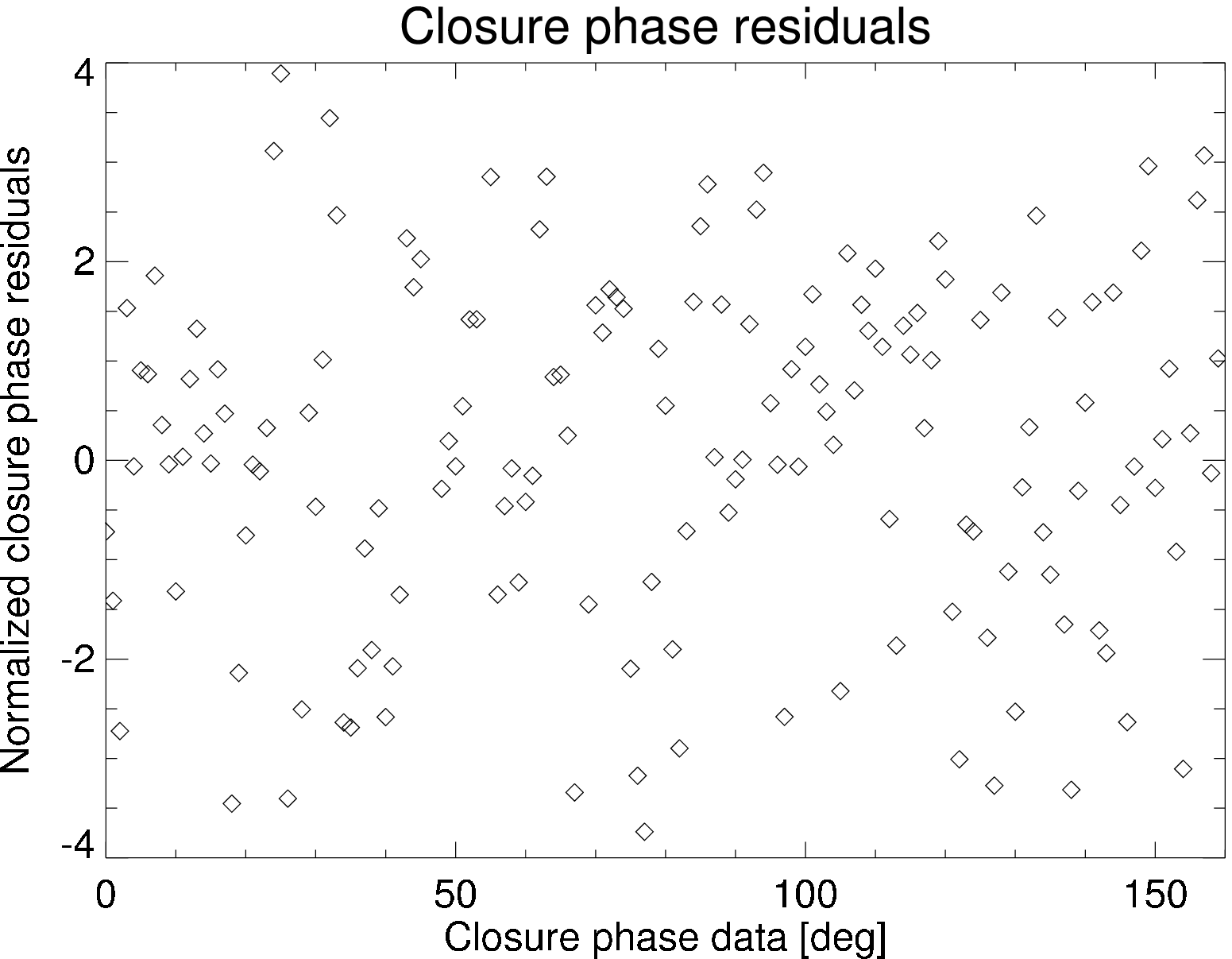}
\caption{Power spectrum residuals (left) and closure phase residuals
  (right) normalized by the data error, here for the 2009 August 12 a
  data set, with a reduced $\chi^2$ of 1.16. \label{fig-modelvsdata}}
\end{figure}

Table~\ref{tbl-3} reports these reduced $\chi^2$ values for every
night. Typically the agreement between the model and the data is
good, especially considering the unrealistic assumption of uniform
brightnesses: the values lie mostly within the 1.0--6.0 range, with
higher $\chi^2$ values encountered during the eclipse epochs where
this model breaks down. Figure~\ref{fig-modelvsdata} presents the
residuals of a typical fit, which show no noticeable trend.

\begin{figure}
\centering
\epsscale{1.0}
\includegraphics[width=0.496\linewidth]{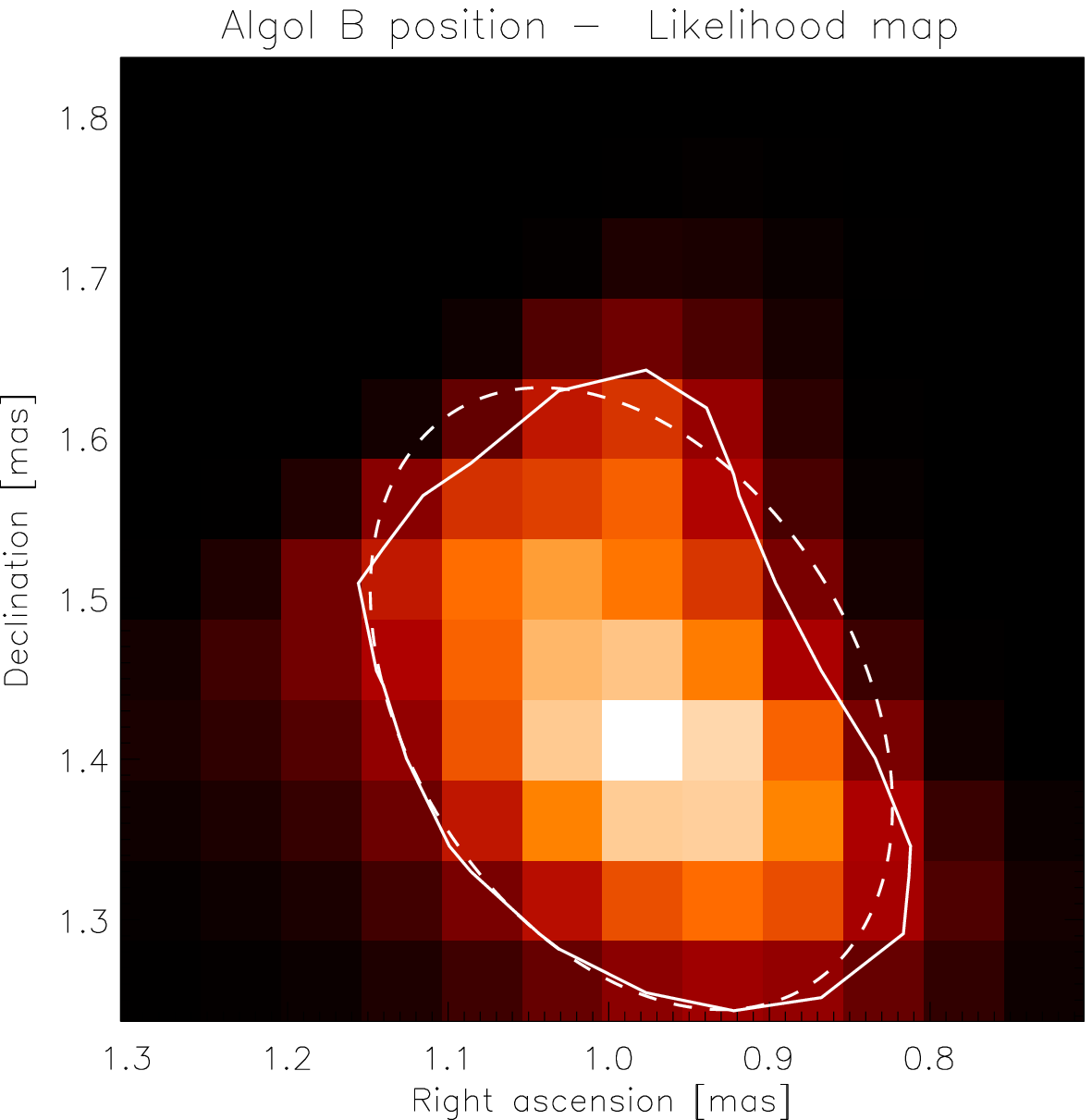}%
\includegraphics[width=0.506\linewidth]{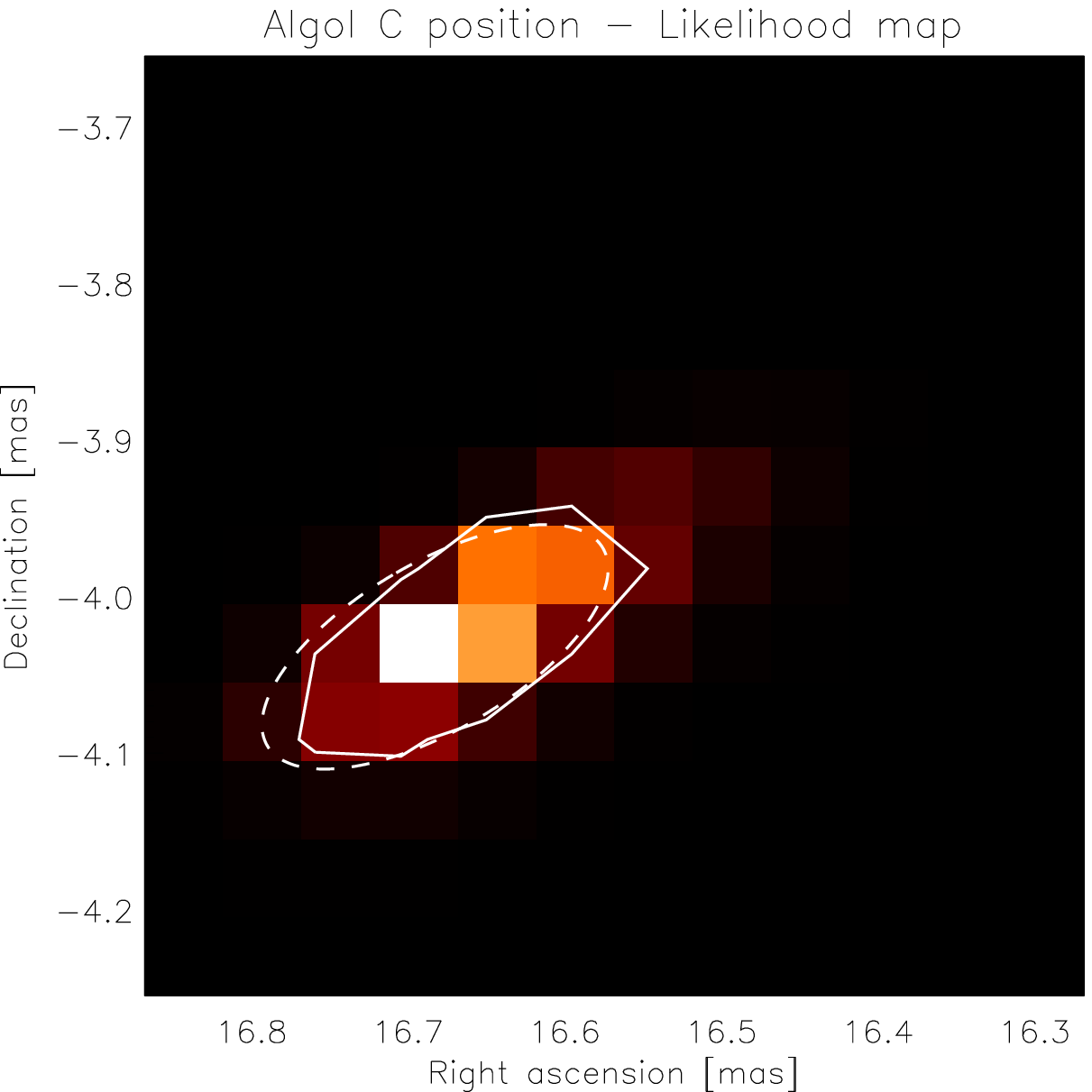}%
\caption{Likelihood maps of Algol~B (left) and Algol~C (right) positions, for the typical 2009 August 12 a
  reconstruction. The solid line corresponds to the 1$\sigma$
  confidence level, and the dashed line is the best-fitting error
  ellipse.\label{fig-chi2maps}}
\end{figure}

The full output of our model-fitting software consists in reduced
$\chi^2$ maps as a function of Algol~B and Algol~C positions, as presented
on Figure~\ref{fig-chi2maps}. The most probable positions of Algol~B and
Algol~C correspond to the peak value of these maps. For each map, the
contour at~1$\sigma$ ($68\%$ confidence level) is computed and fitted
by an ellipse. The positions of the peaks for Algol~B and Algol~C as
well as the parameters of the error ellipses (major and minor axes,
angles) are recorded in Table~\ref{tbl-3}.

The fits also produce estimates for the angular diameters of all three
stars. The estimates are $\phi_A = 0.88 \pm 0.05 $ mas, $\phi_B = 1.12
\pm 0.07 $ mas, and $\phi_C = 0.56 \pm 0.10 $ mas. For Algol~B, this
angular diameter corresponds to the major axis of the ellipse (the
aspect ratio of the Algol~B ellipse changes with the phase of the inner orbit,
between~$1.04$~and~$1.22$ for our data sets, which we attribute the projection
of the Roche lobe).

Concerning Algol~C, we note that the difficulty of estimating the
bandwidth smearing factors is reflected here as a larger error bar
associated with its angular diameter. Our angular diameters are not
directly comparable with the published values in Table~1 of
\citet{Zavala2010}, as these suffered from a conversion error
(i.e. the parallax values in milliarsceconds and parsecs have been
confused). Nevertheless, as they were based on the linear diameters
estimated in~\citet{Richards1993}, we settle for comparison
with~\citet{Richards1993} only. We will assume the most recent value
published for the orbital parallax, $34.7 \pm 0.6$ mas, independently
determined by optical \citep{Zavala2010} and radio
\citep{Peterson2011} interferometry, and more precise than the latest
{\it Hipparcos} estimate of $36.17\pm1.40$ mas
\citep{VanLeeuwen2009}. Finally the linear stellar radii computed with
this parallax are given in Table~\ref{tbl-4}: our results are
consistent with the radius estimates derived by light curve analysis in
\citet{Richards1993}, though they admittedly have greater error bars
due to the night-by-night nature of our fits and the uniform
brightness assumption.

\section{Orbital solution}\label{orbital_solution}

Using the estimates of the relative positions of the components
established in Section~\ref{model_fitting}, the orbital parameters for
the inner and outer orbits are derived.

\begin{figure}
\centering
\includegraphics[width=0.5\linewidth]{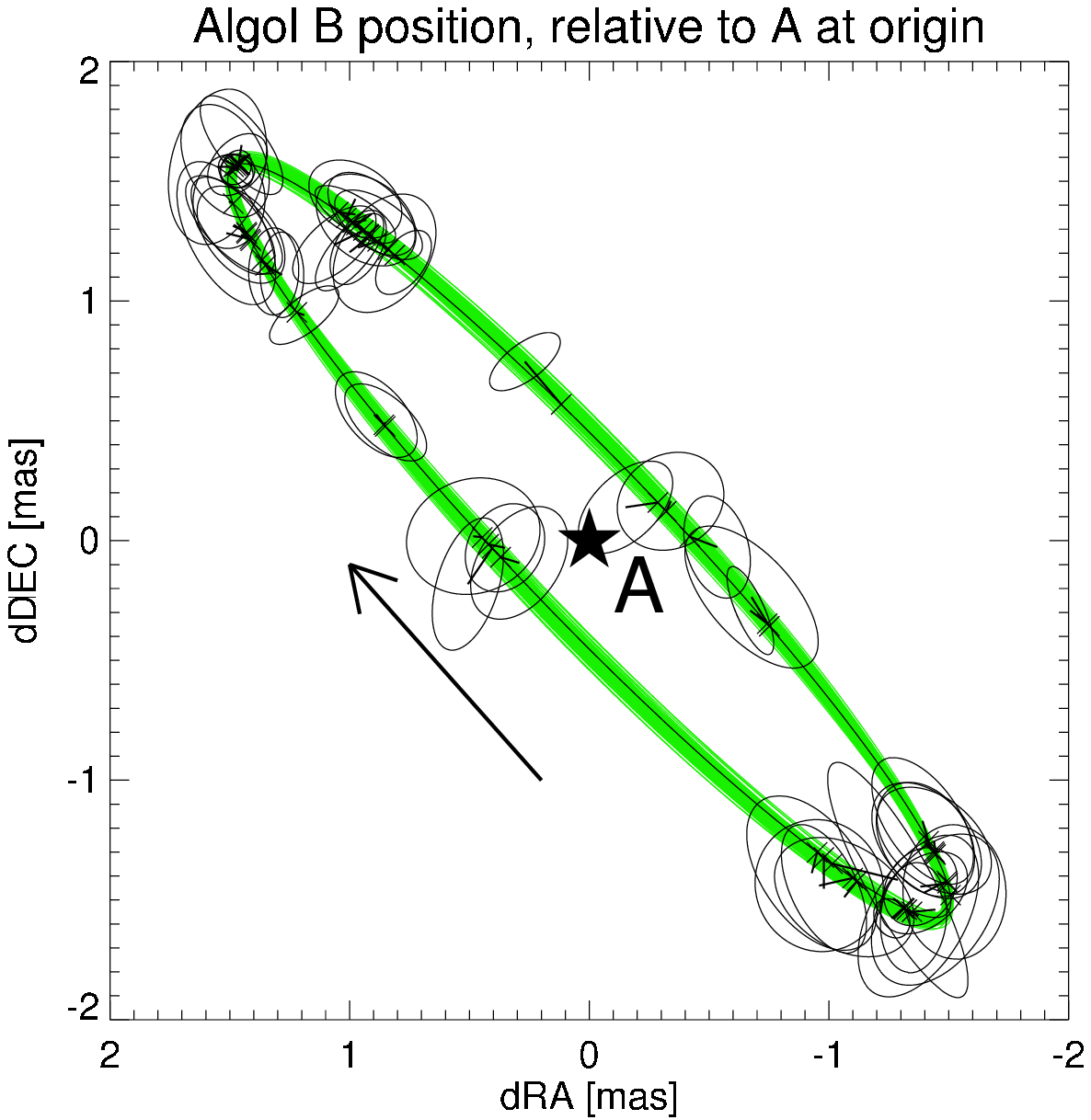}%
\includegraphics[width=0.5\linewidth]{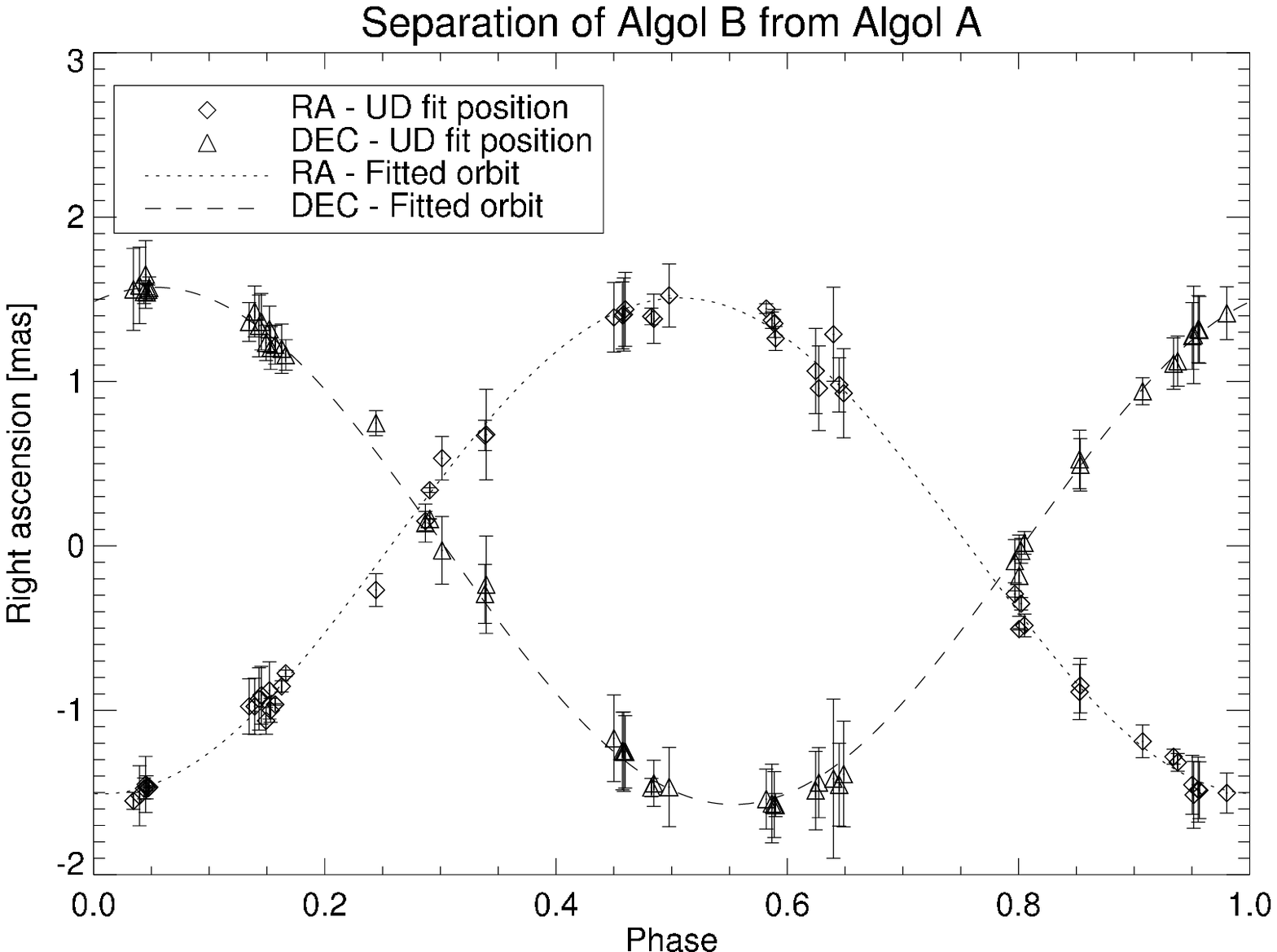}
\caption{Orbit of the inner binary: (left) band of allowed
  orbits at $3\sigma$ for Algol~B relative to Algol~A (green), and
  best fit solution (solid black line) ; the error ellipses for each
  epoch are derived from the $\chi^2$ distribution maps obtained when
  fitting a uniform brightness model; (right) declination and
  right ascension as a function of the inner orbit phase, the zero
  phase corresponding to the light minima.\label{fig-inner_orbit}}
\end{figure}

\begin{figure}
\centering
\includegraphics[width=0.5\linewidth]{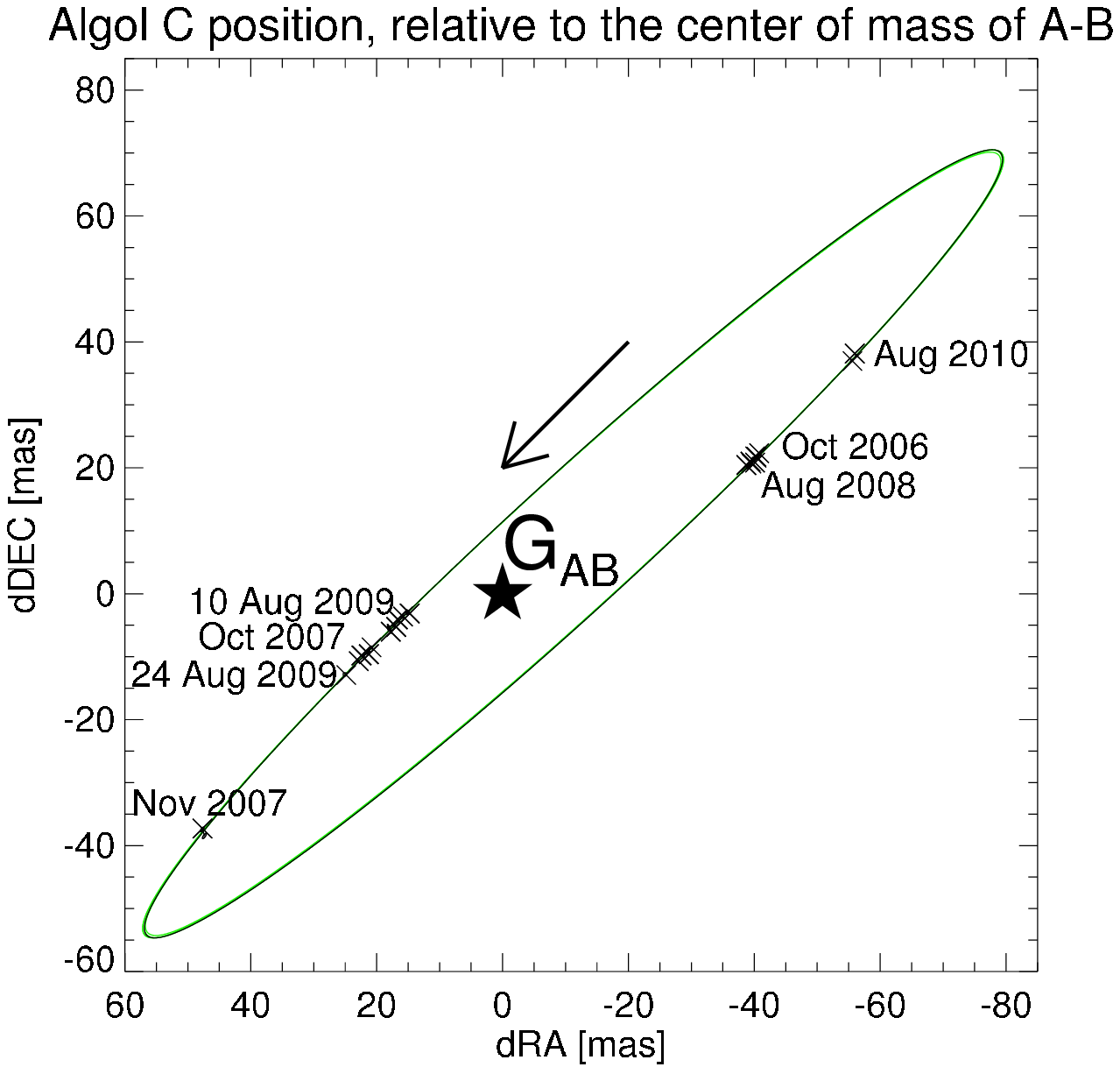}%
\includegraphics[width=0.5\linewidth]{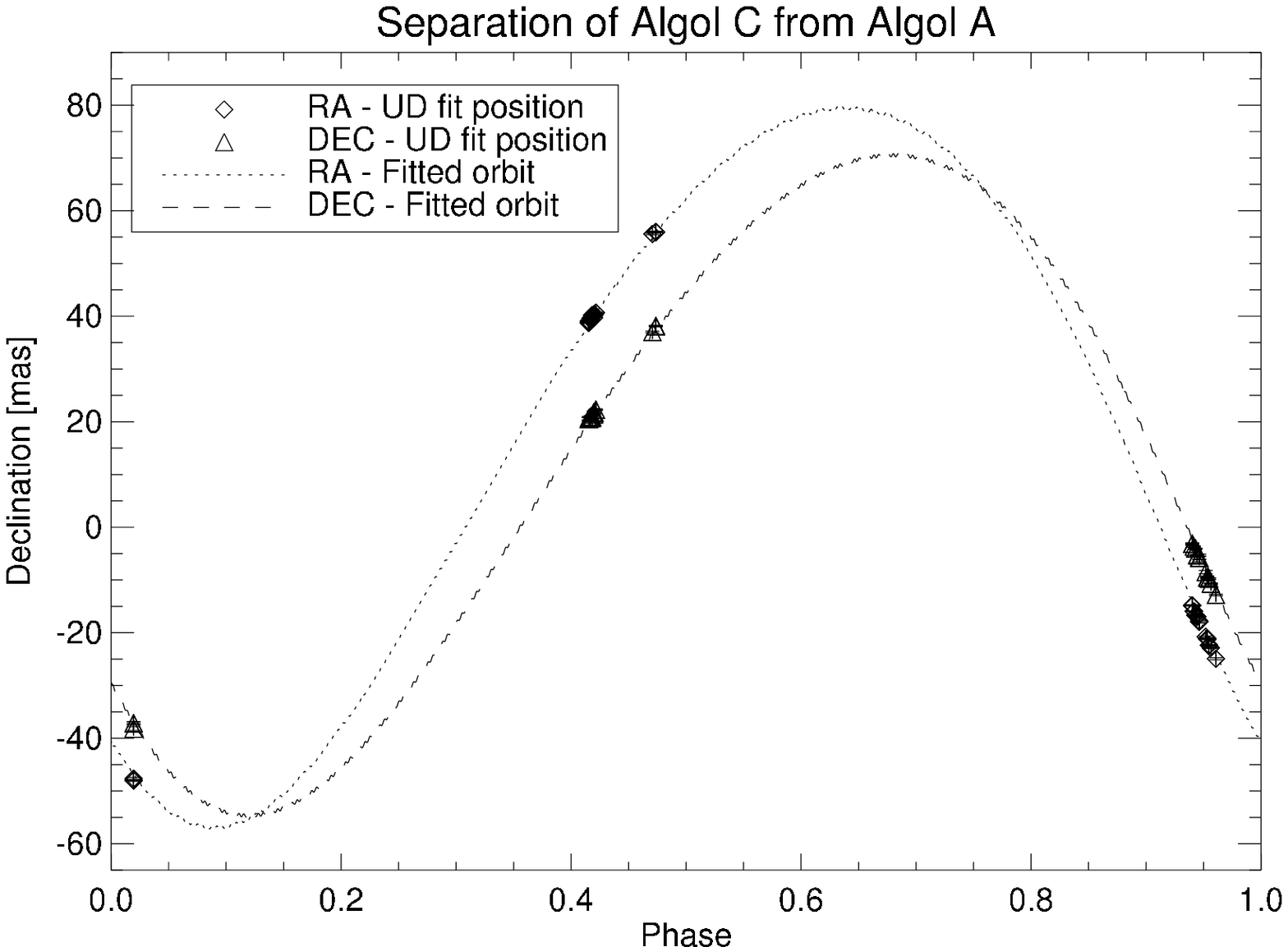}
\caption{Orbit of the outer binary: (left) band of allowed
  orbits at $3$-$\sigma$ for Algol~C relative to the center of mass of
  the inner binary A+B (green), and best fit solution (solid black
  line); (right) declination and right ascension as a function
  of the outer orbit phase (the reference zero phase is at $T_0$ given
  in Table~\ref{tbl-5}); the semi-major axes of the error ellipses
  are at worse $0.5$ mas and thus barely
  visible. \label{fig-outer_orbit}}
\end{figure}

The orbital elements for both orbits and the mass ratio~$m_A/m_B$ are
fitted simultaneously, again using the Levenberg--Marquardt
algorithm. The parameters are initialized around the published values
in \citet{Zavala2010}, for which the corresponding reduced~$\chi^2$ is
about~$13.0$. After convergence our final best~$\chi^2$ is~$0.94$. In
addition to these orbital parameters, Kepler's third law applied to
both the inner and outer orbits allows us to derive the mass ratio
$m_C/(m_A+m_B)$ independently from parallax assumptions. The
parameters corresponding to the best fit are given in
Tables~\ref{tbl-4}--\ref{tbl-6}, with error bars
corresponding to a $68\%$ confidence level. The probability
distributions of the parameters were obtained with the
bootstrapping-with-replacement technique (using $50000$ bootstrap
samples) and thus takes into account correlated errors. In the case
where distributions were found to be roughly symmetrical about the
peak values, only a single error bar is
given. Figures~\ref{fig-inner_orbit} and~\ref{fig-outer_orbit}
show, respectively, the inner and outer orbits corresponding to our best
fit, as well as the estimated positions of Algol~B and Algol~C as
determined in the previous section.

Most of our estimated orbital elements are in agreement with previous
results based on light curve analysis \citep{Richards1993} and radio
and optical interferometry \citep{Zavala2010, Peterson2011}. In
particular the respective prograde and retrograde movements of the
outer and inner orbits are unambiguously verified (see also the
reconstructed image sequence of the inner orbit in the next
section). The mutual inclination angle $i_m$ between both orbits is
given by:
\begin{equation}
\cos i_m = \cos i_1 \cos i_2 + \sin i_1 \sin i_2 \cos (\Omega_1 - \Omega_2)
\end{equation}
where $i_1$ and $i_2$ are the inclination of the inner and outer
orbits, and $\Omega_1$ and $\Omega_2$ are the associated ascending node
angles. The mutual inclination determines the period of orbital
precession between the orbits, and consequently the amplitude of the
variations of inclination of the close binary due to dynamical effects
\citep{Borkovits2004}. Previously published values close to $100
\degree$ \citep{Kiseleva1998} would imply a detectable change of
inclination of the close binary from $1\degree$ to $3\degree$ in the
last century, which in turn would lead to observable variations in the
minimum depth of the light curves of the eclipse and the disappearance
of these eclipses after a few centuries \citep{Csizmadia2009}. This is
however contrary to the actual observations: Algol eclipses have been
known since Antiquity and therefore the minimum depth is not thought
to have significantly changed within error bars. The mutual
inclination estimate derived from our bootstrap results is $i_m =
90\degree.20 \pm 0\degree.32$. It is completely consistent with
photometric observations, and much closer to exact perpendicularity
than the most recent estimate of $95\degree \pm 5\degree$
\citep{Zavala2010}.

\begin{figure}
\epsscale{0.7}\centering
\plotone{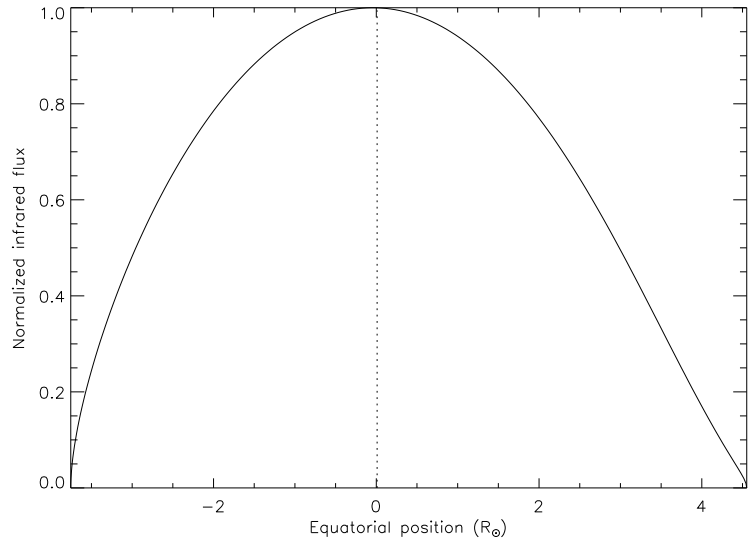}
\caption{Simulated profile of the equatorial brightness of Algol~B for
  a Roche lobe filling factor of~$1.0$. As the surface departs from
  sphericity, the radius-dependent gravity darkening effect induce a
  shift of the photocenter (dashed line) with respect to the center of
  mass (zero position). For this simulation, as we move along the equator, we
  compute the distance of the surface to the center of mass, using the
  Roche-von Zeipel equation. We then derive the corresponding local
  gravity and local temperature values. The brightness distribution is
  given as the combination of gravity darkening and limb darkening. We
  used the results from infrared light-curve fitting published in
  \citet{Richards1990} to set the gravity darkening coefficient
  ($\beta=0.08$, corresponding to the theoretical result for a
  convective envelope) and the quadratic limb-darkening coefficients
  ($u_1= 0.680$, $u_2=-0.171$). The reported brightness distribution
  is normalized to its maximum value. Overall we find that the
  position of the center of light differs from that of the center of
  mass by only $1\%$.\label{fig:roche1d}}
\end{figure}

The main point of disagreement between previous studies and our work
is that we found the semi-major axis of the inner orbit to be slightly
smaller, in turn implying a smaller value for the total mass of the
inner binary. However, to disregard this shorter axis result we would
have to invoke a $15\%$ error in our estimation of the position of the
center of mass of the secondary. As our analysis is essentially based
on the determination of photocenters, a potential source of error to
consider is the asymmetric brightness distribution on the surface of a
Roche-lobe-filling star. We quantified the magnitude of such an
effect, by simulating the equatorial brightness profile of Algol~B,
taking into account both gravity darkening and limb~darkening as shown
in Figure~\ref{fig:roche1d}.  The positions of the photocenter and of
the center of mass were found to coincide to $1\%$. Arguably the
existence of an additional strong proximity effect (due to the primary
heating of the atmosphere of the secondary) could also explain a
photocenter shift. Detailed modeling of such effect however goes
beyond the scope of this paper, requiring complementary data (imaging
in Section~\ref{image_reconstruction} based on current data did not
allow detection of the proximity effect). Finally, the semimajor axis
of the inner binary was also found to be $15\%$ smaller than previous
optical results by recent radio \citep{Peterson2011} and X-ray
observations \citep{Chung2004}, a result strikingly similar to
ours. While both these papers hypothesized the existence of a coronal
component around Algol~B, no coronal signature has been detected in
the near-infrared yet. Our result thus suggests the possibility that
the center of light and center of mass actually roughly coincide, with
previous optical results suffering from imprecisions. Our observations
indeed benefited from much higher angular resolution than previous
interferometric studies and from the recent disambiguation of the
orbital angles from \citet{Zavala2010}.

Consequently we recommend that our orbits (like all previously
published orbits based on interferometric data) should be interpreted
as photocenter orbits. We also note that we did not notice significant
differences between the center of lights estimated by model-fitting
uniform ellipses and estimated by using the real brightness
distribution. In Section~\ref{image_reconstruction} we demonstrate
that the semimajor axes estimated from the disk model and from the
reconstructed images differ by no more than $5\%$.

\begin{figure}
\epsscale{0.7}\centering
\plotone{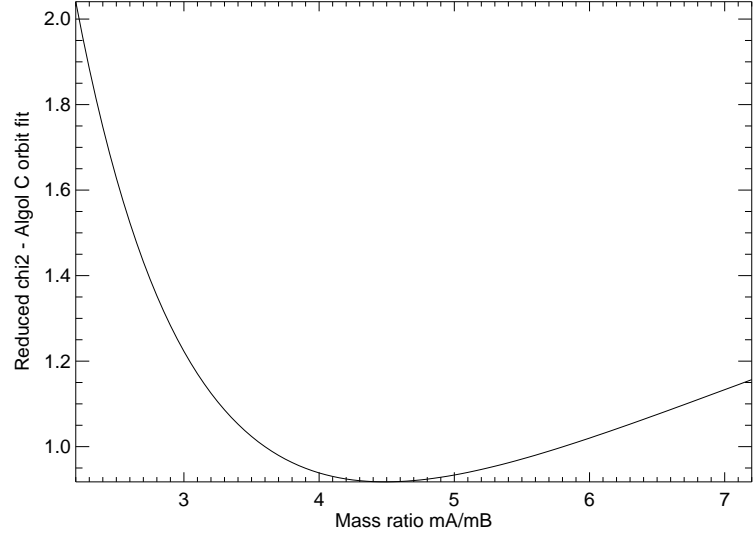}
\caption{Reduced $\chi^2$ of the best fit of Algol~C orbit as a
 function of the inner pair mass ratio, with a minimum at
 $m_A/m_B=4.56 \pm 0.34$. \label{fig-massratio}}
\end{figure}

The inner binary mass ratio~$m_A/m_B=4.56 \pm 0.34$ agrees with the
most recent radial velocity measurements \citep{Hill1993, Hill1971, Tomkin1978}
which measured $m_A/m_B=4.6 \pm 0.1$. Our larger error bar is due to
the lesser intrinsic precision of interferometry in this parameter, as the
mass ratio influences the position of Algol~C only by a small amount.
Figure~\ref{fig-massratio} shows that the $\chi^2$ of the Algol~C
orbit fit possesses a very flat minimum valley, resulting in the large
error on the mass ratio.

With the assumption of a parallax of $34.7\pm0.6$ mas
\citep{Zavala2010, Peterson2011}, the masses of the three stars are
completely determined and we report the results in
Table~\ref{tbl-4}. Algol~A and Algol~B are lighter than found by
\citet{Zavala2010} and \citet{Richards1993}. We note here that the
tertiary mass published in \citet{Zavala2010} is numerically
inconsistent with the corresponding semi-axes and periods, which seems
to be due to slightly different parallax values adopted for the inner
and outer orbit fits (C.~Hummel, private communication, 2011).  With our
hierarchical fit, we find the mass of Algol~C to be larger than that
published in \citet{Richards1993}. Overall our mass estimates are
consistent with all measurements (radial velocity and system dynamics)
and give a sensibly different view of the system. In particular the
picture of the mass transfer within the inner binary may have to be
modified.

\section{Image reconstruction}\label{image_reconstruction}

The amount of data present in each MIRC data sets corresponds to a
dozen minutes of observation each night. All data sets except one
(2009 August 18) contain four-telescope data. Consequently there is
enough phase information to attempt ``model-independent image
reconstruction'' on each night, where the prefix ``model-independent''
underlines that the method does not rely on a specific astrophysical
model.

The image reconstruction procedure belongs to the class of ``ill-posed''
inverse problems, as it attempts to reconstruct an image
conventionally consisting in a large number of pixels (typically a few
thousand) using the less numerous interferometric data points
(typically a few hundred). Expressing the problem in a Bayesian
framework shows that the solution can be given by the regularized
maximum likelihood method \citep{Baron2010, Thiebaut2010}. The target image is the
array of fluxes~$\widehat{\V{x}}=\{x_0, \ldots x_{n-1}\}$ that minimizes
the sum of a term linked to the data (the likelihood) and a term
reflecting all prior information (the regularization) under the two
constraints of image positivity and of normalization of the image to
unity, i.e. in the mathematical form:
\begin{equation}
\widehat{\V{x}} = \underset{\V{x} \in \mathbb{R}^n}{\argmin} \left\{ \chi^2 (\V{x}) + \mu R(\V{x}) \right\} \label{reconst_eq} .
\end{equation}
subject to positivity ($\forall i$, $x_i \ge 0$) and normalization
($\sum_{i=0}^{n-1} x_i = 1$). The likelihood term measures the
distance between the observed power spectrum and bispectrum and the
same quantities derived from the tentative image. Minimizing this term
enforces the presence of flux as indicated by the data, but the
minimization presents many local minima. The addition of the
regularization function $R(\V{x})$ discriminates between these local
minima, as well as prevents over-fitting, detrimental to image
quality. The factor~$\mu$~in Equation~(\ref{reconst_eq}) controls the
relative weight of the $\chi^2$ and regularization terms, and can be
chosen so that the actual reduced $\chi^2$ is roughly unity for the
reconstructed image (the selection of the truly optimal $\mu$ is a
difficult problem which goes beyond the scope of this paper). Most
regularizers are computed in the image plane, but also possesses good
frequency extrapolation properties in the Fourier plane. Hence they
allow reconstructions to routinely achieve super-resolution, i.e., to
produce images with an effective resolution typically about four times
greater than the physical array resolution.

The image reconstruction software SQUEEZE \citep{Baron2010} has been
used to obtain the reconstruction presented in this paper. It combines
the Markov Chain Monte Carlo and gradient-descent approaches
used respectively by its predecessors MACIM \citep{Ireland2006} and
BSMEM \citep{Baron2008a}. SQUEEZE offers a vast choice of
regularizers, ranging from classic entropy to novel wavelet
regularizations. As the images of Algol are expected to consist of
three compact limb-darkened stars, we choose to use the total
variation regularizer, an edge-preserving regularizer which ensures
both the smoothness and the compactness of the stars while strongly
penalizing stray flux~\citep{Rudin1992}. In our implementation, the
total variation is implemented as:
\begin{equation}
R_1(\V{x}) = \sum_i - \operatorname{TV}(x_i) = \sum_i |x_{i+1}-x_i|,
\end{equation}
A second regularizer is also employed in the form of a prior image to
initialize the Markov Chain to a sensible starting point as well as to
prevent the exploration of pixels where the presence of flux is known
to be very improbable. The prior image $\V{m}$ is created by
convolving the best model image of Section~\ref{model_fitting} to one
fourth of the array resolution. The entropic prior expression is the Burg entropy~\citep{Burg1975}:
\begin{equation}
R_2(\V{x}) = \sum_i - \log(x_i/m_i) .
\end{equation}
The number of pixel elements in the Markov Chain is set to~$4000$, and
the length of the Chain to~$10000$ iterations. A common size and
pixellation is chosen for all~$55$ images to allow easier
comparison. The highest instrumental resolution is given by the
largest CHARA baseline (S1-W1, $330$~m), corresponding to $0.5$~mas in
the H~band. Taking into account the expected amount of
super-resolution, our choice of pixellation for the reconstructed
images is thus~$0.1$~mas. Meanwhile the distance from Algol~C to the
inner pair is at most~$100$~mas, thus requiring an image size of $1000
\times 1000$ pixels. This is a factor $100$ more pixels than typical
binary reconstructions from non-simulated data in optical/infrared
interferometry \citep{Zhao2008}. Nevertheless all~$55$~reconstructions
converged without notable issues, with final reduced $\chi^2$ between
$1.0$ and $2.0$.

\begin{figure}
\centering
\plotone{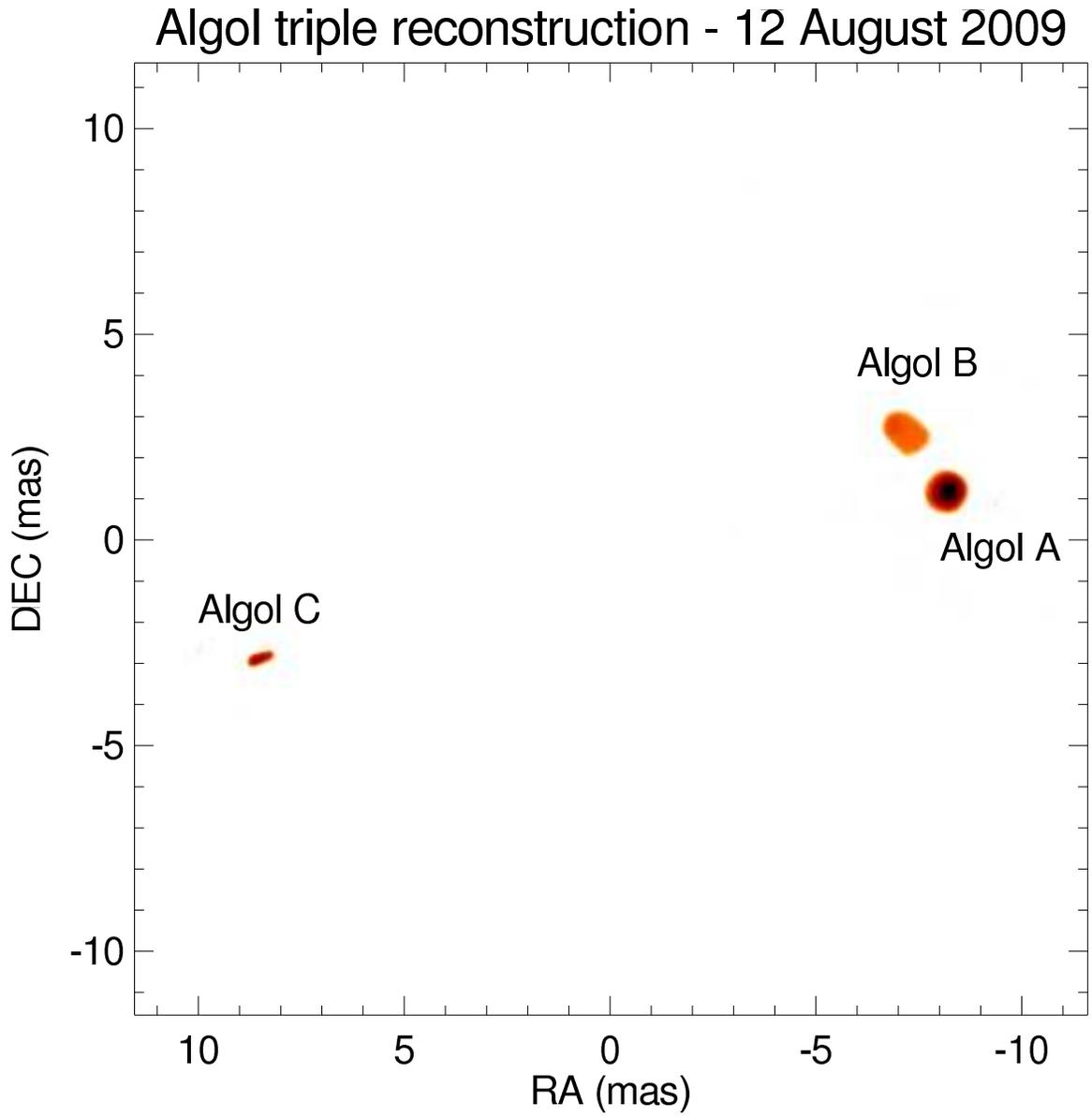}
\caption{Reconstructed image of the triple system for the 2009 August
  12 a data set (final reduced $\chi^2$=1.02). The three components are
  resolved: Algol~A is nearly circular, Algol~B is elongated as it
  fills its Roche lobe, and Algol~C is elongated due to bandwidth
  smearing. \label{fig-image_ABC}}
\end{figure}

Figure~\ref{fig-image_ABC} presents the reconstruction for 2009
August 12. The three stars are present within a small field of 16
mas, Algol~C being close to its periapsis, and they are well resolved.

Algol~A appears as a bright and near-perfect circular disk with a
small amount of limb-darkening. Algol~B is fainter and elongated
toward Algol~A, as expected due to it filling its Roche lobe. Algol~C
also appears elongated in the direction of Algol~A--B, though this is
only here a purely optical-numerical artifact due to bandwidth
smearing.

\begin{figure}
\centering
\includegraphics[width=\linewidth]{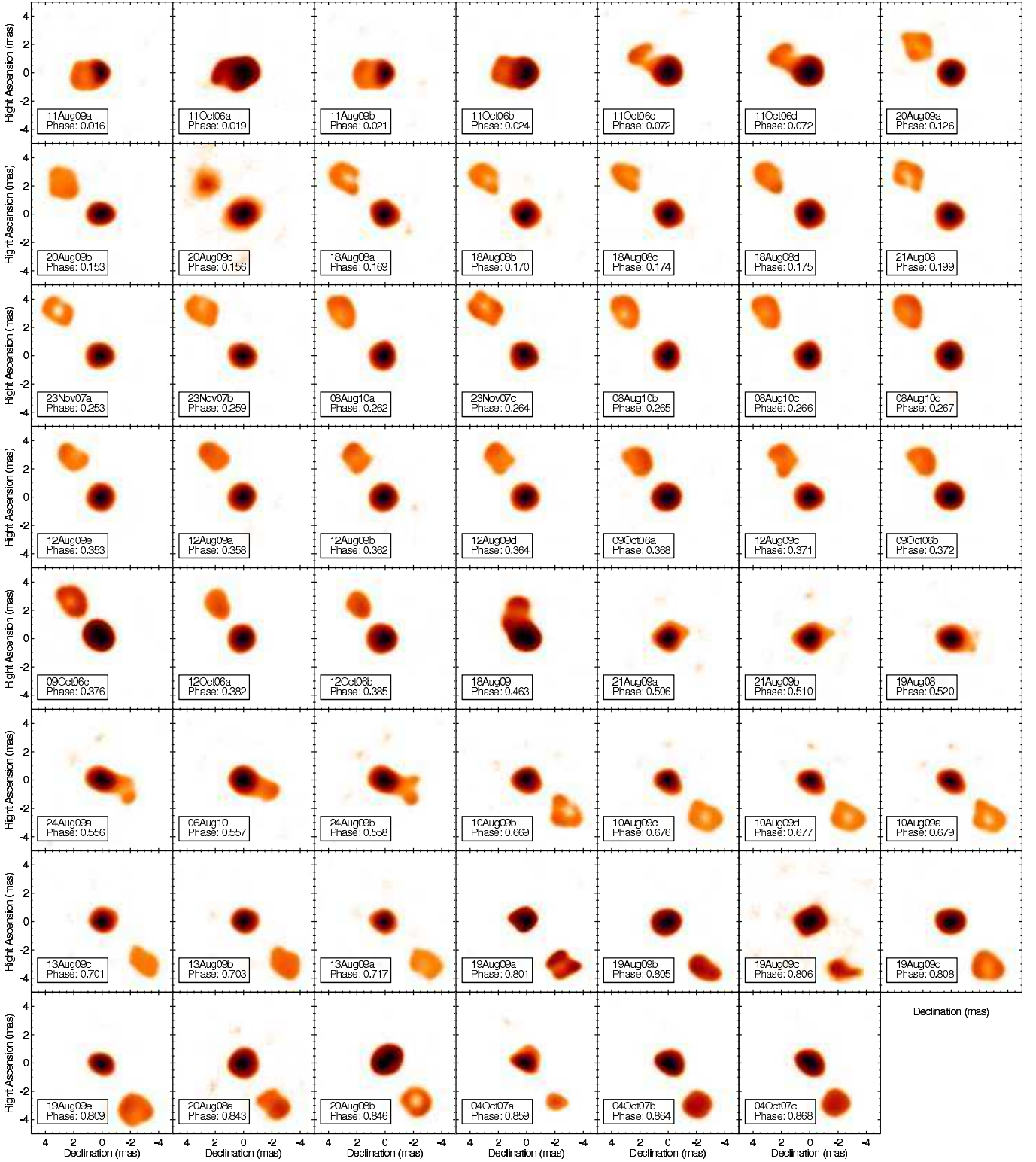}
\caption{Evolution of the SQUEEZE reconstructions of the inner stellar
  system A--B with the phase of the primary eclipse. The phase is given
  as $\Phi = [(T_0 - \text{MJD})/P_{\text{AB}} \mod 1]$ with the orbital
  elements from Table~\ref{tbl-6}; $\Phi=0$ corresponds to the primary
  eclipse and the time of the light minima, and $\Phi=0.5$ corresponds
  to the secondary eclipse. \label{fig-image_reconstructions}}
\end{figure}

All the reconstructed images of the inner pair are presented in
Figure~\ref{fig-image_reconstructions}, with epochs sorted by
increasing phase of the inner eclipse. A movie generated from this
sequence is also available with the online version of this
paper. These images should be interpreted with caution, due to the
comparatively poor resolution of the interferometer compared to the
size of the stars, and to the typically limited dynamic range of image
reconstruction with four telescopes (about 10:1). Overall the aspect
of Algol~A barely changes from image to image, while Algol~B's
elongation varies as expected with the phase of the eclipse.

Due to the sporadic nature of our Algol observations, the phase
coverage is unequal and incomplete, and in particular there are no data
available beyond phase~$0.861$. About 20\% of the images correspond to
snapshots taken during the eclipse epochs. Images with a phase below
$0.08$ correspond to the primary eclipse, when Algol~A is occulted by
Algol~B. The progressive separation of the stars is clear as the phase
increases up to $0.25$ (even though the data for these images were
taken several years apart). Images with a phase between~$0.45$
and~$0.55$ correspond to the secondary eclipse. During part of the
secondary eclipse, Algol~B seems to completely disappears behind
Algol~A despite its larger size, an effect we attribute to its lower
brightness and the low dynamic range.

In a significant number of images the secondary is clearly
limb~brightened, a phenomenon we attribute to gravity darkening. We
can neither rule out nor confirm the presence of spots on the stellar
surfaces, such as hypothesized in \citet{Richards1990, Richards1992}.
Similarly, we are unable to confirm the existence of a stream of
matter between the stars or of an active corona around Algol~B such as
detected in radio (while the 2009~August~18 image at phase~$0.463$
would seem to show some interesting interaction between the stellar
pair, it is the least reliable of all images due to data availability
on three telescopes only).

\begin{figure}
\epsscale{0.7}\centering
\plotone{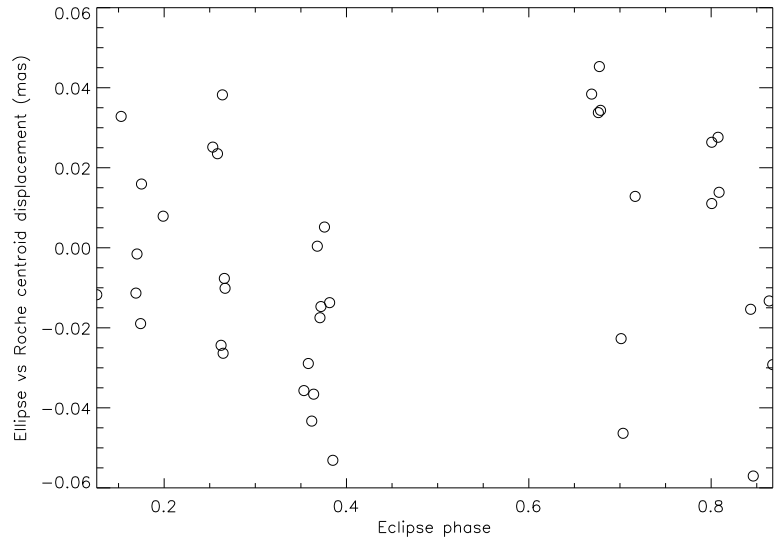}
\caption{Difference between the principal axis of the inner pair
  estimated from the reconstructed images and estimated from uniform
  ellipse model-fitting, as a function of the inner orbit phase. The
  principal axis was defined as the vector joining the centers of
  light of Algol~A and Algol~B, Its computation was restricted to
  epochs where the stellar disks do not
  overlap. \label{fig-uniform_vs_image}}
\end{figure}

Using our reconstructed images, we can now quantify the impact of our
assumption of uniformly bright stars adopted in
Section~\ref{model_fitting}.  As a function of the inner orbit phase,
Figure~\ref{fig-uniform_vs_image} presents the amount of correction to
the principal axis of A--B introduced by using the brightness
distributions given by the imaging. No significant trend is
detected. The maximum correction is $0.057$~mas, roughly corresponding
to half a pixel, and thus always within one standard deviation of the
positions given in Table~\ref{tbl-3}. Consequently, this validates our
modeling in Section~\ref{model_fitting} and increase our confidence in
the derived orbital solution.

\section{Conclusion}

From the analysis of the CHARA/MIRC infrared data on Algol we derived
new orbital elements of the triple system with unprecedented
precision. The respectively prograde and retrograde nature of the
outer and inner orbits is unambiguously confirmed and the mutual
inclination of the orbital planes is found to be extremely close to
perfect perpendicularity. Our results suggest that the semi-major axis
of the inner orbit is shorter than previously reported by other
methods. While this effect could in theory be due to a shift of the
photocenter with respect to the center of mass, we did not find
evidence to support this assumption. If our orbits reflect the center
of mass orbit, then the masses of the primary and secondary are found to be
lighter than suggested by previous works. Using model-independent
image reconstruction techniques, we reconstructed a sequence of 55
images of the inner binary with an effective resolution of 0.2
mas. This ``movie'' of Algol~A and Algol~B covers most phases of the
inner pair, including the primary and secondary eclipses. While the
array resolution did not allow us to confirm the detection of
potential flares or spots on the surface of the stars, the Roche-lobe
shape of the secondary is clearly visible.

To further improve the orbital solution presented in this paper,
three-dimensional and time-dependent image reconstruction algorithms
are currently being developed at the University of Michigan. By
modeling the secondary as a Roche lobe and by imaging directly on
spheroids, we expect to considerably enhance the quality of the
reconstructions, enabling us to detect potential surface features on
the stars.

\acknowledgments

We thank Bob Zavala (NOI) and Christian Hummel (ESO) for their
constructive criticisms on the manuscript.

We are grateful to the reviewers for directing our attention to
the potential effects of the asymmetric brightness distribution on a Roche
lobe surface.

The authors acknowledge funding from the NSF through
awards AST-0807577 to the University of Michigan. 

Operational support for the CHARA Array is provided by the National
Science Foundation through grant AST-0908253 and by the College of
Arts and Sciences at Georgia State University.



\facility{CHARA (MIRC)}








\bibliographystyle{apj}

\clearpage



\begin{deluxetable}{ccrrrrrrrrcrl}
\tabletypesize{\scriptsize}
\tablecaption{Days of observation of Algol with CHARA/MIRC\label{tbl-1}}
\tablewidth{0pt}
\tablehead{\colhead{UT Date} & \colhead{Telescope configuration} & \colhead{Calibrators} }
\startdata
2006 Oct 9	&	S2-E2-W1-W2	& Zet Per \\
2006 Oct 11	&	S2-E2-W1-W2	& Zet Per \\
2006 Oct 12	&	S2-E2-W1-W2	& Zet Per \\
2007 Oct 4 & 	S1-E1-W1-W2 & 37 And, Zet Per \\
2007 Nov 23 & 	S1-E1-W1-W2 & Gam Tri \\
2008 Aug 18 & 	S1-E1-W1-W2 & Zet Per \\
2008 Aug 19 & 	S1-E1-W1-W2 & Zet Per \\
2008 Aug 20 & 	S1-E1-W1-W2 & 37 And \\
2008 Aug 21 & 	S1-E1-W1-W2 & 37 And, Zet Per\\ 
2009 Aug 10	&	S1-E1-W1-W2 & Gam Tri\\
2009 Aug 11	&	S1-E1-W1-W2 & Gam Tri\\
2009 Aug 12	&	S1-E1-W1-W2 & Gam Tri\\
2009 Aug 13	&	S1-E1-W1-W2 & Gam Tri\\
2009 Aug 18	&	S1-W1-W2& Gam Tri\\
2009 Aug 19	&	S1-E1-W1-W2 & Gam Tri\\
2009 Aug 20	&	S2-E2-W1-W2 & Gam Tri, Zet Per\\
2009 Aug 21	&	S2-E2-W1-W2 & Gam Tri \\
2009 Aug 24	&	S2-E2-W1-W2 & 37 And \\
2010 Aug 6 & 	S2-E1-W1-W2 & Gam Tri, 10 Aur\\
2010 Aug 8 & 	S2-E1-W1-W2 & Gam Tri, 10 Aur \\
\enddata
\end{deluxetable}

\begin{deluxetable}{cccc}
\tabletypesize{\scriptsize}
\tablecaption{Calibrator sizes and~1$\sigma$ Error in Milli-arcseconds.\label{tbl-2}}
\tablewidth{0pt}
\tablehead{\colhead{Calibrator} & \colhead{Uniform Disk Size (mas)} & \colhead{Error (mas)} & \colhead{Reference} }
\startdata
10 Aur & 0.419 & 0.063 & b\\
37 And & 0.682 & 0.030 & a,b\\
Gam Tri & 0.522 & 0.033 & a, b, c\\
Zet Per & 0.703 & 0.021 & b \\
\enddata
\tablerefs{(a) \citet{Kervella2008} ; (b) \citet{Barnes1978} ; (c) \citet{Bonneau2006}.}
\end{deluxetable}

\clearpage

\clearpage

\begin{deluxetable}{lcccccccccccc}
\tabletypesize{\scriptsize}
\tablecaption{Position estimates of Algol~B and Algol~C relative to Algol~A\label{tbl-3}}
\tablewidth{0pt}
\tablehead{\colhead{Observation}&\colhead{MJD}&\colhead{$\chi^2/\text{dof}$}&\colhead{$\rho_B$}&\colhead{$\theta_B$}&\colhead{$\sigma_{a, B}$}&\colhead{$\sigma_{b,B}$}&\colhead{$\psi_{B}$}&\colhead{$\rho_C$}&\colhead{$\theta_C$}&\colhead{$\sigma_{a, C}$}&\colhead{$\sigma_{b, C}$}&\colhead{$\psi_C$} \\ \colhead{(UT date)}&\colhead{(days)}&\colhead{}&\colhead{(mas)}&\colhead{(deg)}&\colhead{(mas)}&\colhead{(mas)}&\colhead{(mas)}&\colhead{(mas)}&\colhead{(deg)}&\colhead{(mas)}&\colhead{(mas)}&\colhead{(deg)}}
\startdata
2006 Oct 9 a & 54017.535773 & 1.452 & 1.630 &   40.74 & 0.214 & 0.103 &  318.16 &   44.54 &  --61.91 & 0.217 & 0.140 &  302.82 \\
2006 Oct 9 b & 54017.547162 & 1.193 & 1.565 &   39.72 & 0.199 & 0.083 &  310.04 &   44.55 &  --61.93 & 0.192 & 0.141 &  299.08 \\
2006 Oct 9 c & 54017.558304 & 2.778 & 1.558 &   38.24 & 0.170 & 0.094 &  299.65 &   44.52 &  --61.88 & 0.229 & 0.158 &  106.26 \\
2006 Oct 11 a & 54019.403382 & 6.467 & 0.538 &  109.70 & 0.286 & 0.129 &  286.92 &   45.09 &  --61.73 & 0.331 & 0.202 &  108.53 \\
2006 Oct 11 b & 54019.416411 & 4.271 & 0.485 &   87.79 & 0.283 & 0.238 &  160.59 &   45.62 &  --61.73 & 0.421 & 0.179 &  305.31 \\
2006 Oct 11 c & 54019.553074 & 3.673 & 1.033 &   59.38 & 0.217 & 0.118 &  227.40 &   45.74 &  --61.66 & 0.243 & 0.139 &  119.75 \\
2006 Oct 11 d & 54019.555134 & 3.188 & 0.983 &   59.86 & 0.211 & 0.103 &  222.20 &   45.88 &  --61.72 & 0.234 & 0.150 &  298.48 \\
2006 Oct 12 a & 54020.441845 & 1.124 & 1.473 &   35.46 & 0.271 & 0.183 &  303.77 &   46.37 &  --61.28 & 0.309 & 0.184 &  128.15 \\
2006 Oct 12 b & 54020.451949 & 1.540 & 1.396 &   33.72 & 0.155 & 0.086 &  307.40 &   46.35 &  --61.31 & 0.238 & 0.153 &  136.16 \\
2007 Oct 4 a & 54377.358738 & 2.891 & 1.914 & --137.73 & 0.555 & 0.134 &   61.12 &   18.86 &  108.01 & 0.374 & 0.075 &   67.59 \\
2007 Oct 4 b & 54377.372951 & 2.098 & 1.752 & --146.02 & 0.289 & 0.139 &   64.24 &   18.81 &  107.87 & 0.330 & 0.100 &  242.73 \\
2007 Oct 4 c & 54377.384093 & 2.076 & 1.670 & --146.21 & 0.367 & 0.207 &  234.45 &   18.69 &  108.25 & 0.806 & 0.137 &   64.15 \\
2007 Nov 23 a & 54427.234904 & 3.484 & 2.200 &   44.83 & 0.330 & 0.195 &  103.13 &   61.06 &  128.69 & 0.508 & 0.157 &  124.32 \\
2007 Nov 23 b & 54427.249729 & 1.975 & 2.196 &   43.80 & 0.252 & 0.159 &  242.64 &   60.69 &  127.74 & 0.449 & 0.149 &  125.02 \\
2007 Nov 23 c & 54427.265206 & 2.795 & 2.198 &   41.31 & 0.248 & 0.100 &  232.43 &   60.74 &  127.75 & 0.254 & 0.150 &  309.56 \\
2008 Aug 18 a & 54696.522101 & 5.493 & 1.935 &   48.69 & 0.234 & 0.139 &  231.63 &   44.17 &  --62.15 & 0.344 & 0.109 &  298.54 \\
2008 Aug 18 b & 54696.525448 & 3.876 & 1.985 &   49.72 & 0.326 & 0.158 &  242.68 &   43.80 &  --62.15 & 0.405 & 0.178 &  301.15 \\
2008 Aug 18 c & 54696.536763 & 4.791 & 1.984 &   48.36 & 0.262 & 0.115 &  226.02 &   43.99 &  --62.12 & 0.427 & 0.121 &  293.89 \\
2008 Aug 18 d & 54696.539548 & 3.268 & 1.984 &   48.52 & 0.243 & 0.110 &  231.84 &   43.83 &  --62.15 & 0.401 & 0.121 &  296.55 \\
2008 Aug 19 & 54697.529045 & 3.965 & 0.534 &  --92.89 & 0.219 & 0.130 &  253.86 &   44.67 &  --62.38 & 0.303 & 0.122 &  116.32 \\
2008 Aug 20 a & 54698.455342 & 4.235 & 1.829 & --144.44 & 0.318 & 0.214 &  212.86 &   44.90 &  --62.52 & 0.448 & 0.174 &  128.07 \\
2008 Aug 20 b & 54698.463525 & 1.703 & 1.730 & --146.33 & 0.278 & 0.264 &   62.05 &   45.32 &  --62.46 & 0.471 & 0.151 &  127.81 \\
2008 Aug 21 & 54699.475045 & 3.292 & 2.065 &   46.72 & 0.196 & 0.149 &  273.80 &   45.59 &  --61.46 & 0.299 & 0.086 &  305.89 \\
2009 Aug 10 a & 55053.531881 & 4.257 & 1.908 & --131.04 & 0.276 & 0.175 &  220.41 &   15.17 &  102.16 & 0.320 & 0.186 &  131.00 \\
2009 Aug 10 b & 55053.503913 & 3.033 & 1.818 & --130.08 & 0.302 & 0.153 &  235.73 &   15.15 &  102.10 & 0.223 & 0.177 &  318.71 \\
2009 Aug 10 c & 55053.525093 & 4.020 & 1.877 & --131.65 & 0.258 & 0.179 &  237.61 &   15.16 &  102.12 & 0.315 & 0.180 &  122.35 \\
2009 Aug 10 d & 55053.528275 & 4.348 & 1.884 & --131.65 & 0.279 & 0.182 &  229.05 &   15.17 &  102.12 & 0.325 & 0.172 &  130.78 \\
2009 Aug 11 a & 55054.497896 & 3.595 & 0.307 &  107.63 & 0.261 & 0.167 &  124.70 &   16.35 &  102.97 & 0.269 & 0.194 &  118.47 \\
2009 Aug 11 b & 55054.513533 & 7.212 & 0.353 &   94.87 & 0.194 & 0.153 &  121.59 &   16.38 &  102.91 & 0.208 & 0.148 &  310.20 \\
2009 Aug 12 a & 55055.480222 & 1.157 & 1.728 &   34.39 & 0.210 & 0.141 &  211.54 &   17.18 &  103.62 & 0.124 & 0.053 &  300.49 \\
2009 Aug 12 b & 55055.490959 & 0.724 & 1.630 &   34.83 & 0.231 & 0.163 &  217.99 &   17.18 &  103.68 & 0.191 & 0.155 &  115.96 \\
2009 Aug 12 c & 55055.516855 & 1.169 & 1.585 &   33.70 & 0.211 & 0.100 &  214.40 &   17.20 &  103.76 & 0.114 & 0.091 &  289.18 \\
2009 Aug 12 d & 55055.497004 & 0.983 & 1.639 &   33.68 & 0.219 & 0.123 &  222.33 &   17.19 &  103.69 & 0.099 & 0.093 &  332.47 \\
2009 Aug 12 e & 55055.466371 & 0.578 & 1.676 &   35.63 & 0.194 & 0.071 &   32.54 &   17.19 &  103.65 & 0.115 & 0.062 &  123.01 \\
2009 Aug 13 a & 55056.508177 & 3.446 & 2.115 & --133.91 & 0.279 & 0.215 &  267.54 &   17.79 &  107.45 & 0.266 & 0.188 &  122.84 \\
2009 Aug 13 b & 55056.470177 & 2.552 & 1.999 & --136.27 & 0.175 & 0.138 &  --51.76 &   17.71 &  107.21 & 0.181 & 0.103 &  299.71 \\
2009 Aug 13 c & 55056.463975 & 1.953 & 2.027 & --136.49 & 0.207 & 0.137 &  140.28 &   17.77 &  107.42 & 0.153 & 0.099 &  137.45 \\
2009 Aug 18 & 55061.515703 & 2.238 & 0.794 &   19.77 & 0.175 & 0.076 &  143.17 &   22.44 &  112.29 & 0.360 & 0.062 &  246.93 \\
2009 Aug 19 a & 55062.483659 & 5.451 & 2.111 & --136.85 & 0.349 & 0.219 &  124.05 &   23.14 &  114.34 & 0.432 & 0.179 &  126.58 \\
2009 Aug 19 b & 55062.497488 & 2.599 & 2.083 & --138.78 & 0.369 & 0.200 &  119.47 &   23.21 &  114.27 & 0.336 & 0.195 &  129.47 \\
2009 Aug 19 d & 55062.503528 & 3.903 & 2.076 & --139.32 & 0.229 & 0.137 &  275.62 &   23.29 &  114.34 & 0.190 & 0.114 &  302.48 \\
2009 Aug 19 e & 55062.506739 & 7.770 & 2.020 & --141.33 & 0.087 & 0.068 &   54.47 &   23.22 &  114.25 & 0.099 & 0.061 &  122.80 \\
2009 Aug 20 a & 55063.416862 & 2.399 & 1.516 &   51.64 & 0.174 & 0.072 &  141.62 &   24.38 &  113.59 & 0.195 & 0.129 &  101.55 \\
2009 Aug 20 b & 55063.493563 & 1.571 & 1.696 &   49.13 & 0.178 & 0.090 &   89.27 &   24.44 &  113.54 & 0.191 & 0.118 &  239.39 \\
2009 Aug 20 c & 55063.503760 & 1.862 & 1.731 &   49.53 & 0.182 & 0.104 &   90.96 &   24.44 &  113.48 & 0.180 & 0.162 &   68.83 \\
2009 Aug 21 a & 55064.505563 & 2.316 & 0.206 &  --47.38 & 0.240 & 0.134 &  316.11 &   25.26 &  115.08 & 0.183 & 0.142 &  106.54 \\
2009 Aug 21 b & 55064.516370 & 4.145 & 0.377 &  --64.16 & 0.231 & 0.195 &  330.11 &   25.26 &  115.07 & 0.171 & 0.131 &  199.17 \\
2009 Aug 24 b & 55067.523787 & 8.119 & 0.717 & --109.20 & 0.367 & 0.175 &  227.88 &   28.05 &  117.26 & 0.285 & 0.262 &  291.42 \\
2010 Aug 6 & 55414.467228 & 1.507 & 0.733 & --113.48 & 0.199 & 0.059 &   65.96 &   66.78 &  --56.36 & 0.260 & 0.085 &  133.57 \\
2010 Aug 8 a & 55416.489153 & 0.693 & 2.135 &   43.70 & 0.079 & 0.062 &  403.26 &   67.73 &  --55.73 & 0.139 & 0.053 &  128.93 \\
2010 Aug 8 b & 55416.495590 & 0.737 & 2.130 &   43.59 & 0.068 & 0.073 &   64.08 &   67.73 &  --55.74 & 0.111 & 0.050 &  126.83 \\
2010 Aug 8 c & 55416.499486 & 0.821 & 2.144 &   43.41 & 0.076 & 0.060 &  120.88 &   67.69 &  --55.75 & 0.137 & 0.047 &  125.75 \\
2010 Aug 8 d & 55416.501757 & 0.628 & 2.149 &   43.08 & 0.149 & 0.100 &  312.13 &   67.72 &  --55.75 & 0.222 & 0.079 &  132.75 \\
\enddata
\tablecomments{All angles are given east of north. $\rho_B$ and $\theta_B$ are the polar coordinates of Algol~B, while $\sigma_{a, B}$, $\sigma_{b, B}$, and $\psi_B$ are, respectively, the semimajor axis, semiminor axis, and angular inclination of its error ellipse.}
\end{deluxetable}

\begin{deluxetable}{cccc}
\tabletypesize{\scriptsize}
\tablecaption{Radius and Mass Estimates of Algol~A, Algol~B and Algol~C.\label{tbl-4}}
\tablewidth{0pt}
\tablehead{\colhead{} & \colhead{ Richards et al. (1993)} & \colhead{ Zavala et al. (2010) } & \colhead{ This Work\tablenotemark{a}} }
\startdata
$R_A$ ($R_\odot$) & $2.90 \pm 0.04$ & $\ldots$ & $2.73 \pm 0.20$\\
$R_B$ ($R_\odot$) & $3.5 \pm 0.1$ & $\ldots$ & $3.48 \pm 0.28$ \\
$R_C$ ($R_\odot$) & $1.7$ & $\ldots$ & $1.73 \pm 0.33$ \\
\hline
$M_A$ ($M_\odot$) & $3.7 \pm 0.3$ & $3.7 \pm 0.2$\tablenotemark{b} &$3.17 \pm 0.21$ \\
$M_B$ ($M_\odot$) &$0.81 \pm 0.05$ & $0.8 \pm 0.1$\tablenotemark{b} & $0.70 \pm 0.08$ \\
$M_C$ ($M_\odot$) &$1.6 \pm 0.1$ &$1.5 \pm 0.1$\tablenotemark{b} & $1.76 \pm 0.15$ \\
\enddata
\tablenotetext{a}{Assuming a parallax of $34.7 \pm 0.6$ mas \citep{Zavala2010, Peterson2011}.}
\tablenotetext{b}{Initially based on \citet{Richards1993}, then re-optimized using the Navy Optical Interferometer data.}
\end{deluxetable}

\begin{deluxetable}{cccc}
\tabletypesize{\scriptsize}
\tablecaption{Orbital Parameters for the Outer Orbit.\label{tbl-5}}
\tablewidth{0pt}
\tablehead{\colhead{Orbital Element} & \colhead{Zavala et al. (2010)} &\colhead{This Work} }
\startdata
$T_0$ (JD) & $2446931.6 \pm 0.1$ & $2446927.22^{+ 0.60}_{-0.54}$ \\
$P$ (days) & $679.85 \pm 0.04$ & $680.168 \pm 0.046$ \\
$a$ (mas) & $93.8 \pm 0.2$ & $93.43^{+0.12}_{-0.09}$ \\
$i$ (deg) & $83.7 \pm 0.1$ & $83.66 \pm 0.03$ \\
$\Omega$ (deg) & $ 132.7 \pm 0.1$ & $132.66 \pm 0.08$\\
$e$ & $0.225 \pm 0.005$ & $0.227 \pm 0.002$ \\
$\omega$ (deg) & $310.8 \pm 0.1$ & $310.02 \pm 0.26$ \\
$m_A/m_B$ & \ldots & $4.56 \pm 0.34$ \\
$mC/(mA+mB)$ & \ldots & $0.456 \pm 0.022$ \\
\enddata
\end{deluxetable}

\begin{deluxetable}{ccc}
\tabletypesize{\scriptsize}
\tablecaption{Orbital Parameters for the Inner Orbit\label{tbl-6}}
\tablewidth{0pt}
\tablehead{\colhead{Orbital Element} & \colhead{Zavala et al. (2010)} &\colhead{This work} }
\startdata
$T_0$ (JD) \tablenotemark{a}& $2441771.3395$ & $2441771.353 \pm 0.007$ \\
$P$ (days) & $2.867328$ & $2.867328 \pm 5\times 10^{-5}$ \\
$a$ (mas) & $2.3 \pm 0.1$ & $2.15 \pm 0.05$ \\
$i$ (deg) & $98.6$ & $98.70 \pm 0.65 $ \\
$\Omega$ (deg) & $47.4 \pm 5.2$ & $43.43 \pm 0.32$ \\
$e$ & $0$ & $0$ \\
$\omega$ (deg) & \ldots & \ldots\\
\enddata
\tablenotetext{a}{Time of minimum light at the primary eclipse.}
\end{deluxetable}




\end{document}